\documentclass[aps,twocolumn,showpacs,superscriptaddress,nobalancelastpage,groupedaddress,prb]{revtex4}
\usepackage{fancyhdr}
\usepackage{indentfirst}
\usepackage{graphicx}
\usepackage{amssymb}
\usepackage{amsfonts}
\usepackage{amsmath}
\usepackage{eucal}
\usepackage{eqalign}
\usepackage{plain}
\usepackage{subfigure}
\usepackage{bbm}
\usepackage{bm}
\usepackage{units}
\usepackage{feynmp}
\usepackage{multirow}
\usepackage{tabularx}
\usepackage{array}
\usepackage{exscale}

\newcommand{\mel}[3]{\langle #1 | #2 | #3 \rangle }
\newcommand{\ket}[1]{| #1 \rangle }

\newcommand{\trace}{\mathop{\rm tr}\nolimits}

\newcommand{\Fp}{F_n^{(+)}}
\newcommand{\Bs}{B_n^{(s)}}
\newcommand{\Fm}{F_n^{(-)}}
\newcommand{\Ba}{B_n^{(a)}}

\begin{document}

\title{\textbf{Spin-boson dynamics beyond conventional perturbation theories}}

\author{Francesco Nesi }
\affiliation{Theoretische Physik, Universit\"at Regensburg, 93040 Regensburg,
Germany}
\author{Elisabetta Paladino} \affiliation{MATIS INFM-CNR \& Dipartimento di Metodologie Fisiche e Chimiche,
Universit\`a di Catania, 95125 Catania, Italy}
\author{Michael Thorwart} \affiliation{Institut f\"ur Theoretische
Physik, Heinrich-Heine-Universit\"at D\"usseldorf, 40225  D\"usseldorf, Germany }
\author{Milena Grifoni}
\affiliation{Theoretische Physik, Universit\"at Regensburg, 93040 Regensburg,
Germany}
\date{\today}

\allowdisplaybreaks
\begin{abstract}
A novel approximation scheme is proposed to describe the dynamics of the spin-boson problem. Being nonperturbative in the coupling strength nor in the tunneling frequency, it gives reliable results over a wide regime of temperatures and coupling strength to the thermal environment for a large class of bath spectral densities.
 We use a path-integral approach and start from the exact solution for the two-level system population difference in the form of a generalized master equation (GME).
Then, we approximate inter-blip and blip-sojourns interactions up to linear order, while retaining all intra-blip correlations to find the kernels entering the GME in analytical form. 
Our approximation scheme, which we call Weakly-Interacting Blip Approximation (WIBA), fully agrees with conventional perturbative approximations in the tunneling matrix element (Non-Interacting Blip Approximation) or in the system-bath coupling strength. 
\end{abstract}


\maketitle

\section{Introduction}

Many physical and chemical two-level systems (TLSs) suffer the influence of external environments which cause decoherence effects 
\cite{Leggett87,Weiss99,PhysRep98}. Electron and proton
transfer reaction in condensed phases \cite{Garg85,Bell}, tunneling
phenomena in condensed matter physics \cite{SQUID,Golding92}, or
two-level atoms in an optical cavity \cite{optical} are some very well
known situations susceptible to lack of coherence.
Dissipation processes are to be understood first, in order to provide, if required, further decoherence control schemes.
An important footstep to allow experimental investigation of decoherence mechanisms was the realization of micrometer-sized objects like the radio-frequency superconducting quantum interference device (rf-SQUID) \cite{SQUID}. Recently, due to their scalability and to the ease to experimentally control their parameters, other superconducting devices as the flux-qubit \cite{Chiorescu03} or the Cooper-pair box \cite{Nakamura99,Vion02} have become 
 attractive systems to explore quantum coherence on a fundamental basis and as possible basic unit (\emph{quantum}-bit or \textit{qubit}), for future quantum computers \cite{Makhlin01}. 

 The spin-boson model, where the TLS is bilinearly coupled to a bath
of harmonic oscillators, is 
commonly used to quantitatively describe some aspects of the dissipative dynamics of the above mentioned systems. 
In the spin-boson problem, it is common to evaluate the
dynamics of the expectation value of the pseudo-spin operator
$\langle{\sigma_z}(t)\rangle \equiv P(t)$, as this is the quantity of interest in the
experiments.

To date, the spin-boson model has been mostly described within two main approximation schemes, 
each describing different regimes of temperature and coupling strength to the thermal bath. 
One main road of approximation is based on an expansion to leading order in the tunneling frequency $\Delta$ (see \eqref{spin-boson} below), yielding the so-called noninteracting-blip approximation (NIBA) \cite{Leggett87,Weiss99,note}.
Equations of motion for the dynamical quantity $P(t)$ equivalent to the NIBA were also obtained using projection operator techniques \cite{Aslangul86,Dakhnovskii,Goychuk,Haake,Grabert}. 
The NIBA equations are easily solved numerically and yield analytical solutions in special cases \cite{Leggett87,Weiss99,PhysRep98}. However, for a biased TLS they can be only used in the regime of high temperatures and/or strong friction. An improvement to the NIBA has been performed for a super-Ohmic bath, in the regime of strong coupling \cite{Würger}. 
A more refined approach, which yields nonconvolutive dynamical equations, is the so-called interacting-blip chain approximation IBCA \cite{Winterstetter97,note2}.

In the range of weak TLS-bath coupling and low-temperature, however, the above mentioned schemes fail to properly describe the dynamics of a biased TLS, e.g. a symmetry breaking at zero temperature even for vanishing asymmetries is predicted. In this regime, path-integral methods \cite{Görlich88,Grifoni99,Görlich89} as
well as perturbation theories \cite{Argyres64,Loss05} are used. The lowest order perturbation theory and the path-integral approach
have been shown to  yield the same dynamics for weak
Ohmic damping \cite{Hartmann00}.


To date, only numerical techniques \cite{Egger,Makarov95,Stockburger,Goorden04,FlowEqs97} can
provide a description of the TLS dynamics, being capable to smoothly match between the high and the weak-coupling regimes.
However, ab-initio calculations can become costly if the regime of low temperatures or the long time dynamics are investigated.

In this work we propose 
a novel approximation scheme, which we call 
\textit{weakly-interacting blip approximation} (WIBA), capable to bridge between the weak-coupling and strong-coupling theories. It is based on the observation that bath-induced correlations among ``blips'' and between ``blips'' and ``sojourns'' are intrinsically weak. Thus, in the WIBA, those correlations are included up to first order only. 
As in the NIBA, the time evolution of the population difference $P(t)$ is given by the general master
equation (GME) ($t \ge 0$)
\begin{equation}
\dot{P}(t)=-\int_{0}^t dt'[
K^{a}(t-t')-W(t-t')+K^{s}(t-t')P(t')]\;,
\label{GME1}
\end{equation}
where the irreducible WIBA kernels $K^{s}$ and $K^{a}$ are, respectively, symmetric and antisymmetric with respect to an external bias $\varepsilon$.
As discussed below, such kernels are neither perturbative
in the tunneling matrix $\Delta$ nor in the TLS-bath coupling, see
Eqs. \eqref{sigmasft}, \eqref{sigmaa0ft} and \eqref{wft}. 

Since the inter-blip and blip-sojourn correlations become negligible at high temperatures or large system-bath coupling, the WIBA well matches the NIBA predictions in this regime. On the other hand, at low temperatures and couplings such correlations, neglected in the NIBA, become essential to properly describe the dynamics. In this limit, WIBA perfectly agrees with the predictions of the weak-coupling theories. Finally, in the intermediate coupling and temperature regime where the perturbative approaches fail, the WIBA yields a good agreement with predictions of \textit{ab-initio} calculations.

The paper is organized as follows: In Sec. \ref{sec.model} we introduce the spin-boson model. Then, in Sec. \ref{pathint} the Feynman real-time path-integral formalism is shortly reviewed, as it constitutes the starting point for the approximation schemes discussed in Secs. \ref{eniba} and \ref{s.wiba}. Specifically, we introduce the so-called Feynman-Vernon influence functional and discuss an exact series expression for the expectation value of the system ``coordinate'' $P(t)$.
Since the exact expression is nontrivial, the following Sections are dedicated to some approximation schemes. In Sec. \ref{eniba}, an extended version of the familiar non-interacting blip approximation (\textit{extended}-NIBA) is presented. There we derive a prescription to calculate the kernels entering an equation of the form \eqref{GME1}. 
Sec. \ref{s.wiba} finally contains the major findings of our work. There we introduce the WIBA, which is able to bridge between the weak and strong coupling regimes 
within the same theory. The WIBA kernels entering \eqref{GME1} are found in analytical form. A comparison with predictions of \textit{ab-initio} Quasi-Adiabatic Path-Integral (QUAPI) calculations shows that the WIBA covers a wide spectrum of parameters and constitutes an interpolation between the NIBA and weak-coupling approximation schemes. 

\section{The spin-boson model}
\label{sec.model}

In this Section, the spin-boson model is shortly reviewed, and the dynamical variables of interest are defined.
To start, we consider the pseudo-spin Hamiltonian in the "localized basis" $\{
\ket{L}, \ket{R}\}$,
\begin{equation}
\hat
{\bf H}_0=\frac{\hbar}{2}[\varepsilon\hat\sigma_z-\Delta\hat\sigma_x]\,,
\label{spin-boson}
\end{equation}
with $\hat\sigma_z$ and $\hat\sigma_x$ the Pauli matrices and $\hbar\Delta$ the
energy separation of the two levels at zero bias ($\varepsilon =0$)
which accounts for the \emph{tunneling} dynamics.
Thus, we can interpret $\hat\sigma_z$ as a "position" operator such
that $\mel{L}{\sigma_z}{L} = -1$ and $\mel{R}{\sigma_z}{R} = +1$.
%


%
We choose to model the environment as an ensemble of harmonic
oscillators \cite{Leggett87,Weiss99,Leggett8187} 
which linearly couple to the system "position coordinate"
$\sigma_z$, ending up with a bath Hamiltonian ${\bf H}_{\rm B}$ of
the form (including also the interaction between bath and system)
\begin{equation}
{\bf H}_{\rm B}=
\sum_{j=1}^{\cal N} \frac{1}{2}\Big [\frac{{\bf
p}_j^2}{m_j}+m_j\omega_j^2 {\bf x}_j^2 \Big ] -\sum_{j=1}^{\cal N}
c_j \sigma_z {\bf x}_j\, .
 \label{hamilton}
\end{equation}
The whole system is thus described by the well known
spin-boson Hamiltonian ${\bf H}={\bf H}_{\rm 0}+{\bf H}_{\rm
B}$. In the case of a thermal equilibrium bath, its influence on the
system is fully characterized by the spectral density
\begin{equation}
G(\omega) = \frac{\pi}{2}\sum_{j=1}^{\cal N} \frac{c_j^2}{m_j
\omega_j} \delta(\omega - \omega_j),
\end{equation}
which reduces to a continuous spectral density once the number
${\cal N}$ of harmonic oscillators approaches infinity. Throughout
this work, we choose
 spectral densities with power-law behavior at low frequencies, i.e. ($s>0$)

\begin{equation}
G(\omega)=2\delta_s \omega_{\rm{ph}}^{1-s}\omega^s e^{-|\omega|/\omega_c} \;, \label{deltas}
\end{equation}
with $\delta_s$ being a dimensionless coupling parameter,
$\omega_{\rm{ph}}$ a characteristic phonon frequency, and $\omega_c$
the bath cut-off frequency which is taken to be the largest
frequency in the model. Thus, this class encompasses the commonly
considered Ohmic spectrum with exponential cutoff $G(\omega)=2\alpha
\omega e^{-|\omega|/\omega_c}$, with $\alpha=\delta_1$ being the so
called Kondo parameter of the TLS.




In order to describe the system's dynamics, we focus on the time
evolution of the \textit{reduced density matrix} (RDM)
$\rho(t)=\trace_{\rm B} {\bf W}(t)$, which we obtain after tracing
out all bath degrees of freedom from the total density matrix ${\bf
W}(t)$. 
We choose a factorized initial condition,
namely at time $t=0$ the full density operator ${\bf W}(0)$ is
expressed as a product of the initial system density operator
$\rho^{\rm (S)}(0)$ and the canonical bath density operator at
temperature $T$.
The initial preparation requires some attention \cite{Weiss99}. One could distinguish two different preparations, according to the time when the coupling between system and bath is switched on, which we assume to happen at a time $t_0\le 0$. In the first one, which we refer as ``class A'', the bath is in canonical equilibrium and the system is being prepared at time $t_0=0$ in a certain state, e.g. corresponding to $\sigma_z=+1$. Then, the system evolves out of the state before the environment has relaxed to the shifted equilibrium distribution. This initial preparation is the typical situation in electron transfer reactions when a specific electronic donor is suddenly prepared by photoinjection \cite{Coalson}.
The other situation, ``class B'', is when the system has been held for a long time in a certain state, e.g. $\sigma_z=+1$, so that the bath had time enough to thermalize with the system. It corresponds to choose $t_0\to-\infty$. Then, at time $t=0$, the constraint is released and the system evolves with the spin-boson Hamiltonian. This initial preparation is performed, e.g. in rf-SQUID devices by a suitable choice of an external magnetic field. We shall refer ourselves to such initial preparation throughout the work.

The diagonal elements of
the RDM are called \emph{populations}, whereas the off-diagonal
terms \emph{coherences}. 
In particular, we denote with 
\begin{equation} \label{observable}
P_R(t)=\rho_{\sigma_f=1, \sigma_f'=1}(t)\equiv \rho_{R R}(t)
\end{equation}
the probability of finding the system in
the right state at time $t$ if it was prepared in the right state at time $t=0$ as well (with $P_L$ one
denotes the analogous quantity for the left well).

In the spin-boson problem, it is common to evaluate the
dynamics of the expectation value of the pseudo-spin operator
$\langle{\sigma_z}(t)\rangle=\rho_{R R}(t)-\rho_{L L}(t) \equiv P(t)$, namely
the difference of populations in the
localized basis.
This quantity is also the quantity of interest of many experiments. Similar considerations as discussed in this work for $P(t)$ can be done for the off-diagonal elements of the density matrix.


\section{Real-time path-integral approach to the dynamics} \label{pathint}

In this Section we recall the main steps yielding an exact series expression for $P(t)$, obtained within the path-integral approach.

%

\subsection{Influence functional and bath correlation function}

As we already anticipated, we assume a factorized initial condition
at time $t=0$. 
For such initial condition, 
the exact formal solution for the RDM  can be expressed in terms of
a real time double path integral over piecewise constant forward
$\sigma(\tau)$ and backward $\sigma'(\tau)$ spin paths
\cite{Leggett87,Weiss99,PhysRep98} with values $\pm 1$. The effects of the
environment are in an influence functional inducing
 non-local in time correlations between different path  segments.
Upon introducing the linear combinations $\eta(\tau)=[\sigma
(\tau)+\sigma'(\tau)]/2$, and
$\xi(\tau)=[\sigma(\tau)-\sigma'(\tau)]/2$, the population difference reads
\begin{equation}
P(t)=\int{\cal D}\xi{\cal D}\eta {\cal
A}[\xi,\eta] \exp{\{\Phi[\xi,\eta]\}}\;,
\end{equation}
where ${\cal A}$ is the path weight in the absence of the bath
coupling. 
%
\begin{figure}[t]
\includegraphics[width=0.4\textwidth,bb=184 620 429 721,clip=true]{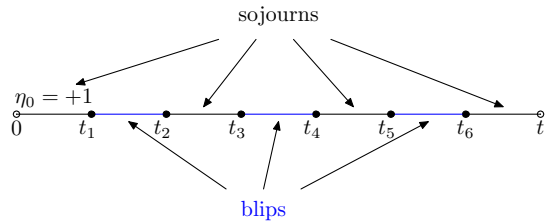}
\caption{Generic path with $2n=6$ transitions at flip times
$t_{1}, t_2, \dots, t_{2n}$. The system is in an off-diagonal state
(blip) of the reduced density matrix (RDM) 
 in the time intervals $t_{2j}-t_{2j-1}$ (blue online) and in a diagonal state (sojourn) during the intervals 
 $t_{2j+1}-t_{2j}$. Due to the initial preparation, the initial sojourn obeys the constraint $\eta_0=+1$.
\label{Fig.freeP}}
\end{figure}
A generic double path can now be visualized as a single path over the four-states of the reduced density matrix, characterized by $(\eta(\tau)=\pm 1\,, \xi(\tau)=0)$ and $(\eta(\tau)=0\,, \xi(\tau)=\pm 1)$. The time intervals spent~in a diagonal $(\xi(\tau)=0)$ and off-diagonal $(\eta(\tau)=0)$ state are dubbed ``sojourns'' and ``blips'', respectively \cite{Leggett87} (see Fig. \eqref{Fig.freeP}).
Due to the initial condition, the path sum runs over all
paths with boundary conditions $\xi(0)=\xi (t)=0$ and $\eta(0)=1$, $\eta(t)=\pm 1$. 
Environmental effects are included in the influence functional
\vspace{-0.1cm}\begin{equation} \label{infl}
\Phi[\xi,\eta] \equiv \int_0^t \!\!dt_2 \int_0^{t_2} \!\!\!dt_1 \,\dot{\xi}(t_2)\left[ S_{2,1} \dot{\xi}(t_1)+ i R_{2,1} \dot{\eta}(t_1) \right]\!\!,
\end{equation}
with the bath correlation function $Q=S+i R$ being
\vspace{-0.11cm}\begin{equation}
Q(t)=\!\!\int_0^\infty \!\!\!\!d\omega
\frac{G(\omega)}{\omega^2}\Big[\coth\Bigl(\frac{\hbar\omega}{2k_{\rm B}T}\Bigr)(1-\cos\omega
t )+i\sin\omega t\Big]
\end{equation}
and $Q_{j,k}:=Q(t_j-t_k)$. 
For the class of spectral densities considered in Eq. \eqref{deltas}, the functions $S$ and $R$ read \cite{Görlich88}
\begin{equation} \label{S}
\begin{split}
 &S(t)=2 \delta_s \Gamma(s-1) \Bigg\{\left(\dfrac{\omega_c}{\omega_{\rm ph}} \right)^{s-1} 
 \bigg[1-\left(1+ (\omega_c t)^2\right)^{\frac{1-s}{2}} \\
&\times \cos{\left[(s-1)\arctan{(\omega_c t)}\right]}\bigg]
+(\hbar \beta \omega_{\rm ph})^{1-s} \bigg[ 2\zeta(s-1,1+\kappa)\\
& - \zeta\left(s-1,1+\kappa+i \frac{t}{\hbar\beta}\right)-\zeta\left(s-1,1+\kappa-i \frac{t}{\hbar\beta}\right) \bigg] \!\Bigg\},
\end{split}
\end{equation}
and
\begin{equation} \label{R}
\begin{split}
 &R(t)=2 \delta_s \Gamma(s-1) \left(\dfrac{\omega_c}{\omega_{\rm ph}} \right)^{s-1} \\
&\times \left(1+ (\omega_c t)^2\right)^{\frac{1-s}{2}} \sin{\left[(s-1)\arctan{(\omega_c t)}\right]},
\end{split}
\end{equation}
where $\Gamma(z)$ is the Euler's gamma function and $\zeta(q,z)$ is the Riemann's generalized zeta function \cite{Gradshteyn65}. Moreover, $\kappa=1/\hbar\beta\omega_c$ becomes important as the ratio $k_{\rm B}T/\hbar \omega_c$ becomes large.

\begin{figure}[t]
\includegraphics[width=0.5\textwidth,bb=100 580 450 730,clip=true]{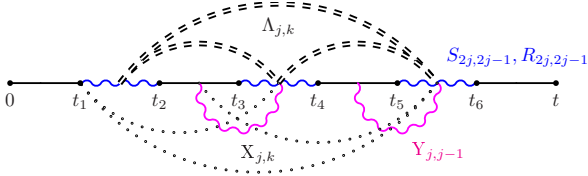}
\vspace{-1.5cm}
\caption{Bath-induced non-local in time correlations among tunneling transitions.
   The interactions $S_{2j,2j-1}$, $R_{2j,2j-1}$ and $Y_{j,j-1}$ (intra-dipole and blip-preceeding-sojourn interactions) which appear in the influence phase $\Phi_{\rm intra, bps}$, cf. Eq. \eqref{influence}, are
symbolized by the wiggled lines (blue and magenta online, respectively). The double-dashed lines denote the
inter-dipole interactions $\Lambda_{j,k}$, while the bold-dotted lines are the remaining
blip-sojourn interactions $X_{j,k}$ contained in the influence phase $\Phi_{\rm inter}$, cf. Eq. \eqref{influencei}.
\label{Fig.interactionsP}}
\end{figure}

%
For a generic path with $2n $ transitions at times $t_j$, $j=1,2,...,2n$, one finds $\dot
\xi(\tau)=\sum_{j=1}^{2n}\xi_j\delta(\tau-t_j)$ and $\dot
\eta(\tau)=\sum_{j=0}^{2n}\eta_j\delta(\tau-t_j)$. Here is $\eta_0=1$
due to the initial preparation and $\xi_j =\pm 1$, $\eta_j=\pm 1$ for $j>0$.
Because $\xi_{2j}=-\xi_{2j-1}$,  the
influence function in \eqref{infl} becomes $\Phi^{(n)}=\Phi_{\rm intra, bps}^{(n)}+\Phi_{\rm
inter}^{(n)}$ (Fig. \ref{Fig.interactionsP}). The function $\Phi_{\rm intra, bps}^{(n)}$ describes intra-blip and blip-preceeding sojourn correlations, and reads
\begin{align}
\begin{split}
\Phi_{\rm intra, bps}^{(n)}& =  - \sum_{j=1}^n\Bigl[S_{2j,2j-1}-i
\xi_j\eta_{j-1}   X_{j,j-1}    \Bigr]\label{influence}\\
& = \Phi_{\rm intra}^{(n)}+ \Phi_{\rm bps}^{(n)}
\;,
\end{split}\\
\Phi_{\rm intra}^{(n)} &=  - \sum_{j=1}^n\Bigl[S_{2j,2j-1}-i
\xi_j\eta_{j-1}   R_{2j,2j-1}    \Bigr]\,, \label{intra}\\
\Phi_{\rm bps}^{(n)} &=  i \sum_{j=1}^n
\xi_j\eta_{j-1}   Y_{j,j-1} \;,   \label{bps}
\end{align}
where we split $X_{j,j-1}= R_{2j,2j-1}+ Y_{j,j-1}$, with 
\begin{align} \label{yjj-1}
 Y_{j,j-1} = R_{2j-1,2j-2}-R_{2j,2j-2}\,.
\end{align}

Moreover, the functional $\Phi_{\rm
inter}^{(n)}$ accounts for inter-blip and blip-sojourns interactions \cite{Leggett87,Weiss99} 
\begin{gather}
\Phi_{\rm inter}^{(n)} =
-\sum_{j=2}^n\sum_{k=1}^{j-1}\xi_j\xi_k\Lambda_{j,k}+i\sum_{j=2}^n
\sum_{k=0}^{j-2}\xi_j\eta_k X_{j,k}\,. \label{influencei}
\end{gather}

The function $\Lambda_{j,k}$  contains the blip-blip interactions
between
 the flip pairs $\{j,k\}$, while the
blip-sojourn  interaction $X_{j,k}$ yields a phase factor.  To be
definite, for $k>0$,
\begin{subequations}
\begin{align}
\hspace{-0.2cm}\Lambda_{j,k}\!&=S_{2j,2k-1}+S_{2j-1,2k}-S_{2j,2k}-S_{2j-1,2k-1}, \label{lint}\\
\hspace{-0.2cm}X_{j,k}\!&=R_{2j,2k+1}+R_{2j-1,2k}-R_{2j,2k}-R_{2j-1,2k+1}. \label{xint}
\end{align}
\end{subequations}

 The correlations $X_{j,0}$  depend on the
 initial preparation, being of class A or B \cite{Weiss99}.

\subsection{Exact series expression for $P(t)$}

 The summation over the path histories reduces to an expansion in the number of tunneling transitions yielding a formally exact series expression for the population difference $P(t)$ \cite{Leggett87,Weiss99}.
It reads (here we identify $\eta_{2n}$ with $\eta_f$) 
\begin{equation}
 P (t)  =  \sum_{\eta_f=\pm 1} \eta_f J(\eta_f,t;\eta_0=1,0) \,,
\end{equation}
with the conditional propagating function being
\begin{equation}
\begin{split}
%
%
 &\hspace{-0.2cm}J(\eta_f,t;\eta_0=1,0)  =  \delta_{\eta_f,\eta_0} + \sum_{n = 1}^\infty \int_0^t \!{\cal D} \{t_j\}\!
\left(-\frac{\Delta^2}{2^2}\right)^n\\
 &\hspace{1.2cm}\times \sum_{\{\xi_j=\pm 1 \}}  B_n \: 
\sum_{\{\eta_j=\pm 1 \}'} \exp{\{\Phi^{(n)}\}}
 \Bigg]\,,
\label{fullpathint2}
\end{split}
\end{equation}
with $B_n \equiv \exp \left\{-{\rm i}
\varepsilon \sum_{j=1}^{n} \xi_j \tau_j \right\}$, and where we defined the blip length as $\tau_j \equiv t_{2j}-t_{2j-1}$. In \eqref{fullpathint2}, $n$ counts the number of blips and the prime in $\{\eta_j=\pm 1 \}'$ means that the sum does not run over the initial and final sojourns, since they are fixed. Moreover,
\begin{equation}
\int_0^t {\cal D} \{t_j\} \equiv \int_0^t dt_{2n} \int_0^{t_{2n}} dt_{2n-1} \dots  \int_0^{t_{2}} dt_1 \;.
\end{equation}

Performing the summation over the intermediate sojourns, one gets for the population difference \cite{Leggett87,Weiss99}:
\begin{equation} \label{pform}
\begin{split}
 P(t)&=1+\sum_{n = 1}^\infty \int_0^t \!{\cal D} \{t_j\}\! \left(-\frac{\Delta^2}{2}\right)^n \\
&\times \sum_{\{\xi_j=\pm 1 \}} \left( \Fp \Bs + \Fm \Ba \right)\,,
\end{split}
\end{equation}
where $\Bs \equiv \cos{\left(\varepsilon \sum_{j=1}^{n} \xi_j \tau_j \right)}$, $\Ba \equiv \sin{\left(\varepsilon \sum_{j=1}^{n} \xi_j \tau_j \right)}$ 
and
\begin{subequations} \label{Fn+-}
\begin{align}
 \Fp &\equiv G_n \prod_{k=0}^{n-1} \cos{\left(\phi_{k,n}\right)},\\
 \Fm &\equiv G_n \sin{\left(\phi_{0,n}\right)} \prod_{k=1}^{n-1} \cos{\left(\phi_{k,n}\right)}\,,
\end{align}
\end{subequations}
with $ \phi_{k,n} \equiv \sum_{j=k+1}^{n} \xi_j X_{j,k}$ and 
\begin{align}
 G_n &\equiv \exp{\left\{\mathfrak{Re}\left[\Phi^{(n)}\right]\right\}} \label{Gn}\\
&= \exp{\left\{ - \sum_{j=1}^n S_{2j,2j-1}
-\sum_{j=2}^n\sum_{k=1}^{j-1}\xi_j\xi_k\Lambda_{j,k}\right\}}\,.
\end{align}

The expression \eqref{pform} is still practically untractable. 
Thus, it is necessary to perform some approximations to describe the TLS dynamics.
Two novel approximation schemes are discussed in the coming Sections \ref{eniba} and \ref{s.wiba}.



\section{The extended Non-Interacting Blip Approximation}
\label{eniba}

\begin{figure}[tb]
\centering
\includegraphics[width=0.6\textwidth,bb=110 611 450 715,clip=true]{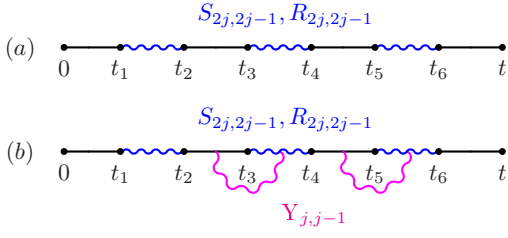}
\vspace{-0.3cm}\caption{Generic path and bath-induced correlations retained in the NIBA (a) and in the \textit{extended}-NIBA (b). In both approximations the inter-blip and blip-sojourns correlations $\Lambda_{j,k}$ and $X_{j,k}$, respectively, which appear in the influence phase $\Phi_{\rm inter}$ (see Eq. \eqref{influence}), are neglected. Within the \textit{extended}-NIBA the blip-preceeding-sojourn interaction $Y_{j,j-1}$ (magenta online), which contributes to the influence phase $\Phi_{\rm bps}$ (see Eq. \eqref{bps}) and is being neglected in the NIBA, is retained. The first sojourn is treated differently, according to the initial preparation (here we choose the preparation of ``class B'').} \label{cfrPniba-ext}
\end{figure}

In this Section we discuss an improvement to the familiar non-interacting blip approximation (NIBA) \cite{Leggett87,Weiss99}, which better treats the blip-preceeding sojourns interactions. We call our more refined approximation scheme ``\textit{extended-}NIBA'' (see Fig. \ref{cfrPniba-ext}).
As we shall see, as the NIBA, the \textit{extended}-NIBA enables to recast the series expression for $P(t)$ into a generalized master equation (GME) of the form \eqref{GME1} with kernels of second order in the level splitting $\Delta$.

%
%
%
%

The advantage of the NIBA relies on its extreme simplicity and on the fact that it is non-perturbative in the coupling to the bath. Hence, the non-interacting blip approximation is a popular approximation scheme. Nevertheless, it has some intrinsic weaknesses, expecially in the asymmetric case (i.e. $\varepsilon\ne 0$) for low temperature and weak coupling.
For example,  NIBA predicts the
unphysical asymptotic limit $\sigma_{z,{\rm N}}^{\infty}=-\tanh
(\frac{\hbar\beta\varepsilon}{2})$, implying a
 localization of the TLS ($\sigma^\infty_{z,{\rm N}}=-1$) at
zero temperature even for vanishing asymmetries. 

The limits of validity of the theory are still dim, this approximation holding whenever the average time spent in a diagonal state (sojourn) $\langle s \rangle$ is very large compared to the average time spent in an off-diagonal state (blip) $\langle \tau \rangle$.
Within the NIBA, the full inter-blip correlations  $\Lambda_{j,k}$
 and the  blip-sojourn interactions  $X_{j,k}$ with $j\ne k+1$ are
 neglected ($\Phi_{\rm inter}^{(n)} \approx 0$). 
The blip-preceeding-sojourn interactions $Y_{j,j-1}$ in Eq. \eqref{yjj-1} are neglected as well (see Fig. \ref{cfrPniba-ext}a). Hence,  
$G_n \approx \exp{\left\{\mathfrak{Re}\left[\Phi_{\rm intra}^{(n)}\right]\right\}}$.
The explicit form of the NIBA kernel is given in Appendix \ref{k.niba}.

In the following Sec. \ref{s.wiba}, we shall introduce a novel approximation scheme, the weakly-interacting blip approximation (WIBA), capable to overcome these drawbacks. Before doing this, however, we need to introduce the \textit{extended}-NIBA where, as in NIBA, $\Phi_{\rm inter}^{(n)} \approx 0$, but the blip-preceeding-sojourn correlations $Y_{j,j-1}$ are retained, despite in approximate form.


\subsection{Series expression within the \textit{extended}-NIBA} \label{GMElam}

After performing the sum over the blip indices $\xi_j=\pm 1$ in Eq. \eqref{pform}, the \textit{extended}-NIBA prescription $\Phi_{\rm inter}^{(n)}=0$ yields for the probability difference $P_{\rm eN}(t)$:
\begin{equation} \label{pniba1}
\begin{split}
 P_{\rm eN}(t)=1 &+\sum_{n = 1}^\infty \left(-1\right)^n \int_0^t \!{\cal D} \{t_j\}\! \left[ g(\tau_1, s_{0}) \right. \\
&\left. + h(\tau_1, s_0) \right] \prod_{j= 2}^n g(\tau_j, s_{j-1}) \,,
\end{split}
\end{equation}
where $\tau_j=t_{2j}-t_{2j-1}$ and $s_j:=t_{2j+1}-t_{2j}$ denotes the sojourn length.
The \textit{extended}-NIBA kernels are defined as
\begin{equation} \label{kniba1}
\begin{split}
g(\tau_j, s_{j-1})&=\Delta^2 {\rm e}^{-S(\tau_j)} \cos(\varepsilon \tau_j) \cos[X_{j,j-1}]\;,\\
h(\tau_j, s_{j-1})&=\Delta^2 {\rm e}^{-S(\tau_j)}
\sin(\varepsilon \tau_j) \sin[X_{j,j-1}]\;.
\end{split}
\end{equation}
Remember that $X_{j,j-1}=R(\tau_j)-R(\tau_j+s_{j-1})+R(s_{j-1})$.
The functions $g(\tau_1, s_{0})$ and $h(\tau_1, s_0)$ explicitly depend on the length $s_0$ of the initial sojourn, and assume a different form for preparation class A and B.

In the following, we choose a factorized initial condition at time $t=0$ with the particle being held at the site $|R\rangle$ ($\sigma_z=+1$) from time $t_0=-\infty$ till $t=0$ (class B), which amounts to consider $s_0 \to \infty$. Hence, Eq. \eqref{kniba1} for $j=1$ reads
\begin{equation} \label{kniba0}
\begin{split}
g(\tau_1, s_{0}\to\infty) &=\Delta^2 {\rm e}^{-S(\tau_1)} \cos(\varepsilon \tau_1) \cos[R(\tau_1)]\equiv {g}_{\rm N}(\tau_1)\;,\\
h(\tau_1, s_{0}\to\infty) &=\Delta^2 {\rm e}^{-S(\tau_1)} \sin(\varepsilon \tau_1) \sin[R(\tau_1)]\equiv {h}_{\rm N}(\tau_1)\;,
\end{split}
\end{equation}
being independent of $s_0$. Notice that $g_{\rm N}(\tau_1)$ and $h_{\rm N}(\tau_1)$ coincide with the symmetric and antisymmetric NIBA kernels, respectively (cf. Eq. \eqref{kniba2}).

Due to the convolutive structure of Eq. \eqref{pniba1}, 
it is easier to evaluate the probability difference upon Laplace transformation. By exchanging the integration order and performing some change of variables, one gets
\begin{equation} \label{pniba2}
\begin{split}
 \hat{P}_{\rm eN}(&\lambda)=\dfrac{1}{\lambda} +\sum_{n = 1}^\infty \left(-1\right)^n \int_0^\infty \!\! {\cal D}_\infty \{\tau_j, s_{j-1}\}   
 \\ 
&\times \left[ g(\tau_1, s_{0}) 
+ h(\tau_1, s_0) \right] \prod_{j= 2}^n g(\tau_j, s_{j-1}) \,,
\end{split}
\end{equation}
where $\hat{P}(\lambda)\equiv \int_0^\infty dt \, \exp{[-\lambda t]} P(t)={\cal L}_t\{P(t)\}$ and 
\begin{equation} \label{intd}
\begin{split}
&\int_0^\infty \!\! {\cal D}_\infty \{\tau_j, s_{j-1}\} \equiv \int_0^\infty \!\! ds_n 
\int_0^\infty \!\! d\tau_n \int_0^\infty \!\! ds_{n-1} \\
&\times 
 \dots \int_0^\infty \!\! ds_1 \int_0^\infty \!\! d\tau_1 \int_0^\infty \!\! ds_0 
\: {\rm e}^{-\lambda(\sum_{j=1}^n (\tau_j+s_{j-1})+s_n)} \\
&=\int_0^\infty \!\! ds_n  {\rm e}^{-\lambda s_n} \\
&\times \int_0^\infty \!\! d\xi_n {\rm e}^{-\lambda \xi_n} \int_0^{\xi_n} \!\! d\tau_n \dots 
\int_0^\infty \!\! d\xi_1 {\rm e}^{-\lambda \xi_1} \int_0^{\xi_1} \!\! d\tau_1 \,,
\end{split}
\end{equation}
with $\xi_j \equiv \tau_j+s_{j-1}$ the length of each blip plus its preceeding sojourn. Let us introduce the functions
\begin{subequations} \label{kFA}
\begin{align}
 F(\xi_j) &\equiv \int_0^{\xi_j} d\tau_j \, g(\tau_j,\xi_j-\tau_j)\,, \quad j>1, \\
 F_0(\xi_1) &\equiv \int_0^{\xi_1} d\tau_1 \, {g}_{\rm N}(\tau_1)\,, \\
 A_0(\xi_1) &\equiv \int_0^{\xi_1} d\tau_1 \, {h}_{\rm N}(\tau_1)\,,
\end{align}
\end{subequations}
noticing that $X_{j,j-1}=R(\tau_j)-R(\xi_j)+R(\xi_j-\tau_j)$. Then $\hat{P}(\lambda)$ assumes the form 
\begin{equation}
\begin{split}
 \hat{P}_{\rm eN}(\lambda)&=\dfrac{1}{\lambda}+\dfrac{1}{\lambda} \sum_{n = 1}^\infty \prod_{j=2}^n 
\left[- {\cal L}_{\xi_j} F(\xi_j)\right](\lambda) \\
&\times \left\{- {\cal L}_{\xi_1} \left[ F_0(\xi_1)+A_0(\xi_1) \right] \right\}(\lambda) ,
\end{split}
\end{equation}
where the $1/\lambda$ comes from the free last sojourn $s_n$, or equivalently 
\begin{align}
 \hat{P}_{\rm eN}(\lambda)
&=\dfrac{1}{\lambda}-\dfrac{1}{\lambda} \sum_{n = 1}^\infty [-\hat{F}(\lambda)]^{n-1} 
[\hat{F}_0(\lambda)+\hat{A}_0(\lambda)] \notag \\
&=\dfrac{1}{\lambda}-\dfrac{1}{\lambda} \dfrac{\hat{F}_0(\lambda)+\hat{A}_0(\lambda)}{1+\hat{F}(\lambda)}. \label{pbfine}
\end{align}
It is convenient to introduce the functions $\hat{K}^{a}_{\rm eN}(\lambda)\equiv \lambda \hat{A}_0(\lambda)$, $\hat{K}^{s}_{\rm eN}(\lambda)\equiv \lambda \hat{F}(\lambda)$ and $\hat{W}_{\rm eN}(\lambda)\equiv \lambda \left[\hat{F}(\lambda)- \hat{F}_0(\lambda)\right]$.
Then Eq. \eqref{pbfine} becomes 
\begin{align}
 \hat{P}_{\rm eN}(\lambda)&= \dfrac{1-\dfrac{\hat{K}^{a}_{\rm eN}(\lambda)-\hat{W}_{\rm eN}(\lambda)} {\lambda}}{\lambda+\hat{K}^{s}_{\rm eN}(\lambda)}\,. \label{pfine}
\end{align}

\subsection{Generalized master equation (GME) for the \textit{extended}-NIBA model} \label{GMEapp}

Eq. \eqref{pfine} can be easily transformed back to the time domain.
We find 
\begin{equation}
\begin{split}
 \dot{P}_{\rm eN}(t)&=-\int_{0}^t dt'[
K^{a}_{\rm eN}(t-t')\\
&-W_{\rm eN}(t-t')+K_{\rm eN}^{s}(t-t')P_{\rm eN}(t')]\;,
\end{split}
\label{GMEen}
\end{equation}
with \textit{extended}-NIBA kernels defined as 
\begin{subequations} \label{deriv}
\begin{align}
K_{\rm eN}^{s} (t)&\equiv {d\over dt} F(t) := \dot{F}(t),\\
K_{\rm eN}^{a} (t) &\equiv {d\over dt} A_0(t) := \dot{A_0}(t),\\
{K}^{s}_{\rm 0,eN} (t)&\equiv {d\over dt} F_0(t) := \dot{F_0}(t),\\
W_{\rm eN} (t) &\equiv K_{\rm eN}^{s} (t)-{K}^{s}_{\rm 0,eN} (t)\,.
\end{align}
\end{subequations}




Although the time derivatives \eqref{deriv} can be straightforwardly calculated from Eqs. \eqref{kFA}, the explicit dependence of the function $g(\tau_j,\xi_j-\tau_j)$ on $\xi_j$ still implies an integral form for the function $\dot{F}(\xi_j)$. An approximate form of $\dot{F}(\xi_j)$ can be obtained if we observe that the average blip length $\langle \tau \rangle$ is suppressed by the intra-blip interaction $\exp{[-S(\tau)]}$ in the integrands of Eq. \eqref{kFA} and that the imaginary part of the bath correlation function $R(\tau)$, Eq. \eqref{R}, slightly deviates from a constant. Hence, we approximate 
\begin{equation} \label{Rtilde}
-R(\xi_j)+R(\xi_j-\tau_j) \approx -\tau_j \dot{R}(\xi_j)+\CMcal{O}[\tau_j^2 \ddot{R}(\xi_j)]\,.
\end{equation}
Corrections proportional to the second derivative of $R(t)$ have been neglected. 
The functions $g(\tau_j, s_j)$ and $h(\tau_j, s_j)$ defined in Eq. \eqref{kniba1} become ($j>1$) 
\begin{equation} \label{gheniba}
\begin{split}
g(\tau_j, \xi_j-\tau_j)&\approx \tilde{g}(\tau_j, \xi_j)\\
&:=\Delta^2 {\rm e}^{-S(\tau_j)} \cos(\varepsilon \tau_j) \cos[R(\tau_j)-\tau_j \dot{R}(\xi_j)]\;,\\
h(\tau_j, \xi_j-\tau_j)&\approx \tilde{h}(\tau_j, \xi_j)\\
&:=\Delta^2 {\rm e}^{-S(\tau_j)} \sin(\varepsilon \tau_j) \sin[R(\tau_j)-\tau_j \dot{R}(\xi_j)]\;,
\end{split}
\end{equation}
the corrections being of order $\CMcal{O}[\tau_j^2 \ddot{R}(\xi_j)]$. 
In particular, we define $g_{\rm eN}(\xi_j)$ and $h_{\rm eN}(\xi_j)$ as
\begin{equation} \label{ghtilde}
\begin{split}
g_{\rm eN}(\xi_j)\equiv \tilde{g}(\xi_j,\xi_j)&=\Delta^2 {\rm e}^{-S(\xi_j)} \cos(\varepsilon \xi_j) \cos[\tilde{R}(\xi_j)]\;,\\
h_{\rm eN}(\xi_j)\equiv \tilde{h}(\xi_j,\xi_j)&=\Delta^2 {\rm e}^{-S(\xi_j)} \sin(\varepsilon \xi_j) \sin[\tilde{R}(\xi_j)]\;,
\end{split}
\end{equation}
where $\tilde{R}(t) \equiv R(t)-t \dot{R}(t)$. 
The extended-NIBA kernels, obtained as prescribed by Eqs. \eqref{deriv} and \eqref{Rtilde}, then read 
\begin{subequations} \label{keniba}
 \begin{align}
 \begin{split}
  K^{s}_{\rm eN}(t) &:= g_{\rm eN}(t) \,, 
 \end{split}\\
 \begin{split}
  K^{a}_{\rm eN}(t) &:= h_{\rm eN}(t) \,, 
 \end{split}\\
 \begin{split}
  {K}^s_{\rm 0,eN}(t)  &:= g_{\rm N}(t) \,,
 \end{split}\\
 \begin{split}
  W_{\rm eN}(t)  &:= g_{\rm eN}(t)-g_{\rm N}(t)  \,.
 \end{split}
\end{align}
 \end{subequations}

\begin{figure}[t]
\centering
\includegraphics[width=0.6\textwidth,bb=90 650 410 700,clip=true]{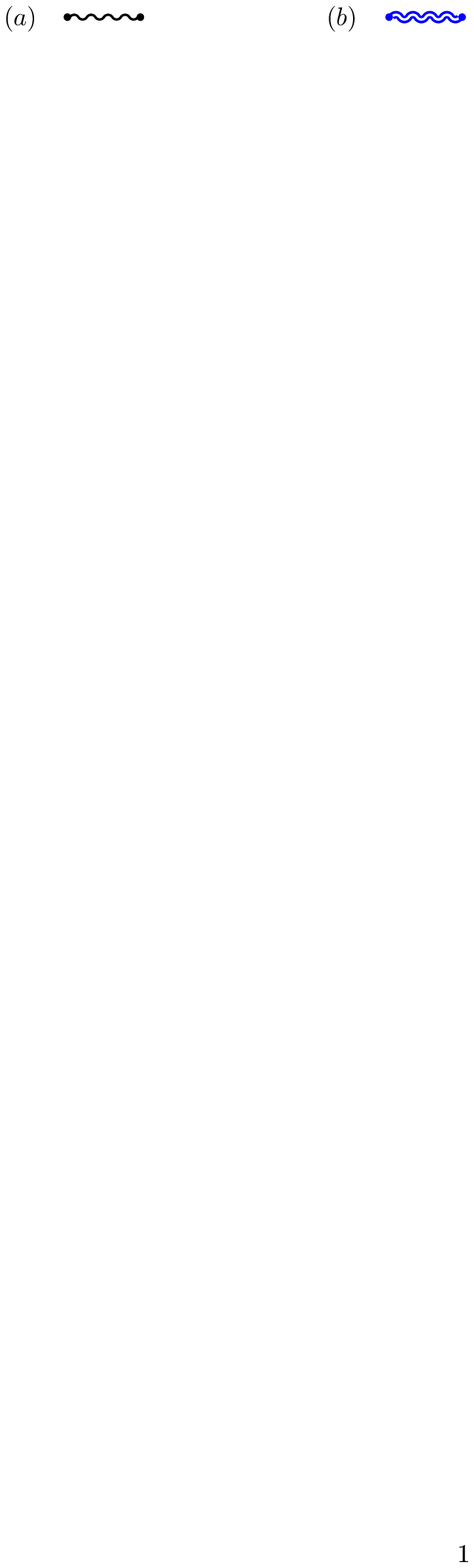}
\vspace{-0.75cm}
\caption{Irreducible kernel $K^{(s)}(t)$ in the
NIBA (a) and in the \textit{extended}-NIBA (blue online) (b). The \textit{extended}-NIBA kernel is obtained from the NIBA one by replacing the function $R(t)$ with $\tilde{R}(t)\equiv R(t)-t\dot{R}(t)$. Hence, the blip becomes ``dressed''.
\vspace{-0.25cm}
\label{Fig.interactions}}
\end{figure}

The irreducible kernel $K^{s}_{\rm eN}(t)$ entering the \textit{extended}-NIBA master equation \eqref{GMEen} is shown in Fig. \ref{Fig.interactions}b. The irreducible NIBA kernel $K^{s}_{\rm N}(t)$, Eq. (\ref{kniba2}a), is depicted in Fig. \ref{Fig.interactions}a. As discussed in App. \ref{k.niba}, within the NIBA an analogous GME as in \eqref{GMEen} is obtained where, due to the approximation $\Phi_{\rm bps}^{(n)}=0$, is $W_{\rm N}(t)=0$. 

\begin{figure}[htb]
\centering
\includegraphics[width=0.45\textwidth,bb=37 56 710 527,clip=true]{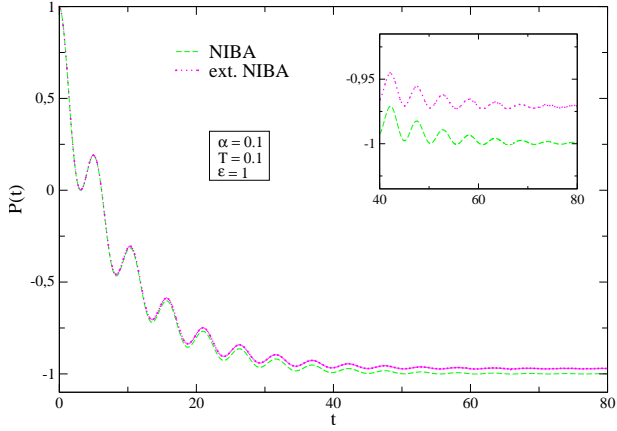}
\vspace{-0cm}\caption{A comparison between the standard NIBA and the \textit{extended}-NIBA for an Ohmic bath ($s=1$) is shown. The parameters are: $T=0.1$, $\alpha=0.1$, $\varepsilon=1$ (in units of the tunneling frequency $\Delta$). One can see that in the chosen intermediate parameter regime the \textit{extended}-NIBA coincides with the conventional one at short times. However, it predicts a different asymptote from the NIBA one (see inset). 
Hence, at moderate damping and temperatures, the blip-preceeding-sojourn correlations retained in the \textit{extended}-NIBA become important.} \label{NIBA-eNIBA}
\end{figure}

The extended-NIBA kernels $K^{s}_{\rm eN}$ and $K^{a}_{\rm eN}$ differ from the conventional NIBA ones, $K^{s}_{\rm N}$ and $K^{a}_{\rm N}$, by the replacing of the imaginary part of the bath correlation function $R(t)$ with the ``dressed'' one $\tilde{R}(t) \equiv R(t)-t \dot{R}(t)$ (see Fig. \ref{Fig.interactions}b). 
At small coupling strength, the NIBA is recovered, since blip-preceeding-sojourn correlations become negligible. 
Despite its simplicity, however, the \textit{extended}-NIBA already yields an improvement to the NIBA in the intermediate coupling regime (see Fig. \ref{NIBA-eNIBA}).

\section{The Weakly-Interacting Blip Approximation (WIBA)}
\label{s.wiba}

\begin{figure}[t]
\centering
\includegraphics[width=0.475\textwidth,bb=55 550 423 723,clip=true]{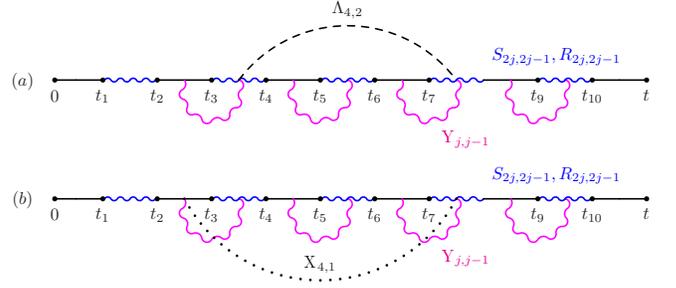}
\vspace{-0.3cm}\caption{Generic paths and bath-induced correlations retained in the WIBA. The linearized influence functional $\Phi_{\rm inter}^{(n)}$ yields one single blip-blip correlation (a) or blip-sojourns interaction (b) for each path.} \vspace{-0.25cm} \label{Pwiba}
\end{figure}

To bridge between the strong damping situation
 described by the NIBA  and the extremely underdamped case we observe
that, for  spectral densities of the form (\ref{deltas}),   the
blip-blip interaction terms $\Lambda_{j,k}$ as well as the
blip-sojourn terms $X_{j,k}$ ($k\neq j-1$) are intrinsically small. Therefore, we propose a novel approximation scheme, which we call
 weakly-interacting blip approximation (WIBA).
Within the WIBA, the full $\Phi_{\rm intra,bps}^{(n)}$ is retained as in the \textit{extended}-NIBA and one expands the influence functional $\exp{\{\Phi_{\rm inter}^{(n)}\}}$ up to linear order in the blip-blip and blip-preceeding sojourns interactions $\Lambda_{j,k}$ and $X_{j,k}$ (see Fig. \ref{Pwiba}).
Hence, 
\begin{equation} \label{wexp}
 \exp{\{\Phi^{(n)}\}}\approx \exp{\{\Phi_{\rm intra,bps}^{(n)}\}} \left(1+\Phi_{\rm inter}^{(n)}\right).
\end{equation}
 In other terms, 
all contributions which involve $\phi_{k,n}$ (cf. Eq. \eqref{Fn+-}) must be expanded up 
to first order in $X_{j,k}$, $j>k+2$, and $G_n \approx \exp{\left\{\mathfrak{Re}\left[\Phi_{\rm intra,bps}^{(n)}\right]\right\}}\left(1+\mathfrak{Re}\left[\Phi_{\rm inter}^{(n)}\right]\right)$ (cf. Eq. \eqref{Gn}). All terms of the order $\Lambda X$ or higher are neglected.
As usual, the first sojourn and blip must be treated differently from the others, according to the initial preparation.

The inter-blip and blip-sojourns correlations $\Lambda_{j,k}$ and $X_{j,k}$ are known to become essential to properly describe the dynamics of a \textit{biased} TLS at low temperatures \cite{Görlich88,Görlich89,Grifoni99}.
Specifically, in Ref.\ \onlinecite{Görlich89} a systematic weak-coupling approximation (WCA) was developed where \textit{all} bath-induced correlations were linearized: $\exp{\{ \Phi^{(n)} \}}\approx 1+\Phi^{(n)}$. It was shown in Ref.\ \onlinecite{Hartmann00} that the WCA exactly matches results obtained within the lowest-order Born approximation for weak-coupling to an Ohmic bath.

Hence, the WIBA constitutes an improvement of the \textit{extended}-NIBA model on one side and of the WCA on the other. 
At high temperatures, where the dipole-dipole interactions are
negligible, the
 WIBA kernels reduce to the \textit{extended}-NIBA ones. By expanding the
 WIBA kernels to first order in
 $\delta_s$,
 the weak damping
  kernels  in Refs.\ \onlinecite{Görlich89,Grifoni99} are recovered.

In contrast to the \textit{extended}-NIBA or the weak-coupling approximation, the WIBA has no small parameter in a strict sense, and it is based on the consideration that the correlations $\Phi_{\rm inter}^{(n)}$ are intrinsically weak over the whole parameter regime. As we show below, the WIBA indeed well describes the TLS dynamics over a wide parameter range.


\subsection{Series expression within the WIBA} \label{GMElw}

Let us again start from Eq. \eqref{pform} and apply the WIBA prescription. 
After performing the sum over $\xi_j=\pm1$ we then find 
\begin{align}
 P_{\rm WIBA}&(t)=1 + \int_0^t dt_2 \int_0^{t_2} \!\!dt_1 [-(g_1+h_1)] \notag\\
&+\sum_{n = 2}^\infty \left(-1\right)^n \int_0^t \!{\cal D} \{t_j\}\! \Bigg\{(g_1+h_1)\left( \prod_{j=2}^n g_j \right) \notag\\
&+\sum_{l=2}^n  \left( \prod_{j=l+1}^n g_j \right) \left[ -\bar{h}_l \left( \prod_{j=2}^{l-1} g_j \right) (\bar{g}_1-\bar{h}_1) \Lambda_{l,1} \right. \notag\\
&+\left.  \bar{h}_l \left( \prod_{j=2}^{l-1} g_j \right) (g_1+h_1) X_{l,0} \right] \notag\\
&+\sum_{l=3}^n \sum_{k=2}^{l-1} \left( \prod_{j=l+1}^n g_j \right) \left[ \bar{h}_l \left( \prod_{j=k+1}^{l-1} g_j \right) 
\bar{h}_k \Lambda_{l,k} \right.  \notag\\
&+\left. \bar{h}_l \left( \prod_{j=k+1}^{l-1} g_j \right) 
h_k  X_{l,k-1}  \right] \left( \prod_{j=2}^{k-1} g_j \right) (g_1+h_1) \Bigg\},
\label{pwiba1}\\
=&P_{\rm eN}(t)+P_{\rm inter}(t)\,,
\end{align}
where $P_{\rm eN}(t)$ is the \textit{extended}-NIBA expression \eqref{pniba1}, while $P_{\rm inter}(t)$ contains the contributions coming from the linearized influence functional $\exp{\{\Phi_{\rm inter}^{(n)}\}}$, cf. \eqref{wexp}.
A graphical representation of the time-ordered sequences entering Eq.\ \eqref{pwiba1} is shown in Fig.\ \ref{Pwiba}.
In \eqref{pwiba1}, $g_j$ and $h_j$ are the functions already occurring in \eqref{pniba1}:
\begin{equation} \label{ghwiba}
\begin{split}
g_j &\equiv g(\tau_j, s_{j-1})\;,\\
h_j &\equiv h(\tau_j, s_{j-1})\;.\\
\end{split}
\end{equation}
The functions  $\bar{g}_j$ and $\bar{h}_j$ are analogously defined as
\begin{equation} \label{ghbwiba}
\begin{split}
\bar{g}_j &\equiv \Delta^2 {\rm e}^{-S(\tau_j)} \cos(\varepsilon \tau_j) \sin[X_{j,j-1}]\;,\\
\bar{h}_j &\equiv \Delta^2 {\rm e}^{-S(\tau_j)} \sin(\varepsilon \tau_j) \cos[X_{j,j-1}]\;.
\end{split}
\end{equation}
Notice that $g_j$ and $\bar{g}_j$ are symmetric in the bias, while $h_j$ and $\bar{h}_j$ are antisymmetric.
Moreover, 
\begin{align} \label{kniba0n}
g_1 & =
{g}_{\rm N}(\tau_1) \;,
\\
h_1 & =
{h}_{\rm N}(\tau_1) \;,
\end{align}
cf. Eq. \eqref{kniba0}, and analogously for $\bar{g}_1$ and $\bar{h}_1$. 
To proceed, let us consider the function (for $k>1$)
\begin{gather}
\begin{split}
 \hspace{-0.5cm}\Sigma_s^{(l-k+1)} &(t_{2l}-t_{2k-2}) \equiv \int_{t_{2k-2}}^{t_{2l}} dt_{2l-1} 
\dots \int_{t_{2k-2}}^{t_{2k}} \!\!dt_{2k-1} \\
&\times \bar{h}_l \left( \prod_{j=k+1}^{l-1} g_j \right) 
\left[ \bar{h}_k \Lambda_{l,k} + h_k  X_{l,k-1}  \right]
\end{split} \label{symm}
\end{gather}
entering \eqref{pwiba1}, where $2(l-k+1)$ denotes the number of tunneling transitions. Upon introducing the variables $\xi\equiv t_{2l}-t_{2k-2}$ and $\xi_j \equiv \tau_j+s_{j-1}$, Eq. \eqref{symm} assumes the form 
\begin{equation}
\begin{split} \label{sigmas}
 \hspace{-0.5cm} \Sigma_s^{(l-k+1)} (\xi) = \int_0^\xi d\xi_k 
\int_0^{\xi-\xi_k} \!\!d\xi_l \gamma_{\bar{h}_l}(\xi_l) f(\xi-\xi_k-\xi_l) \\
\times \left[ \gamma_{h_k}(\xi_k) X_{l,k-1}+ \gamma_{\bar{h}_k}(\xi_k) \Lambda_{l,k}\right]\,,
\end{split}
\end{equation}
where $\gamma_{h_j/\bar{h}_j}(\xi_j)$ is an operator which acts on a generic function $w(\tau_j)$ as (e.g. let us consider the operator $\gamma_{h_j}(\xi_j)$)
\begin{equation} \label{gammah}
 \gamma_{h_j}(\xi_j) w := \int_0^{\xi_j} d\tau_j \, h \,(\tau_j,\xi_j-\tau_j) w(\tau_j)\,.
\end{equation}

In \eqref{sigmas}, the function $f$ is given by the expression 
\begin{align} \label{fs}
\displaystyle &f(t_{2l-2}-t_{2k}) = \delta(t_{2l-2}-t_{2k}) \notag \\
\begin{split}
\displaystyle 
&+ \int_{t_{2k}}^{t_{2l-2}} \!\!dt_{2l-3} \dots \int_{t_{2k}}^{t_{2k+2}} \!\!dt_{2k+1} \left(\prod_{j=k+1}^{l-1} -g_j \right) 
\,.
\end{split}
\end{align}

Upon introducing $s\equiv t_{2l-2}-t_{2k}$ and $\xi_j \equiv \tau_j+s_{j-1}$ as in Eq. \eqref{sigmas}, we find 
\begin{align}
\begin{split}
f&(s)= \delta(s)\\
&+\int_{0}^{s} d\xi_{k+1}  \int_{0}^{\xi_{k+1}} \!\!d\tau_{k+1} \left[ -g(\tau_{k+1}, \xi_{k+1}-\tau_{k+1}) \right] \\
&\times \dots \int_{0}^{s-\sum_{j=k+1}^{l-3} \xi_j} \!\!d\xi_{l-2}  \int_{0}^{\xi_{l-2}} \!\!d\tau_{l-2} \left[ -g(\tau_{l-2}, \xi_{l-2}-\tau_{l-2}) \right] \\
&\times \int_{0}^{s-\sum_{j=k+1}^{l-2} \xi_j} \!\!d\tau_{l-1} \left[ -g\left(\tau_{l-1}, s-\!\!\sum_{j=k+1}^{l-2} \xi_j-\tau_{l-1} \right) \right]
\end{split} \notag \\
&=\delta(s)+\dot{p}_{\rm eN}^{(l-k-1)}(s)\,,
\end{align}
where the first derivative of the conditional probability $p_{\rm eN}$ has been introduced.
It satisfies the master equation for the \textit{extended}-NIBA with the symmetric kernel only:
\begin{equation}
\begin{split}
 \dot{p}_{\rm eN}(t)&=-\int_{0}^t dt'
K_{\rm eN}^{s}(t-t')p_{\rm eN}(t')\;.
\end{split}
\label{GMEenp}
\end{equation}

Analogously, we can treat the contributions to $P_{\rm inter}$ which depend on the initial preparation. Specifically, we introduce for the case $k=1$ (here $\xi\equiv t_{2l}-0$) 
\begin{gather}
\begin{split} \label{sigmaa0}
 \Sigma_{0,a}^{(l)} (\xi) \equiv \int_0^\xi d\xi_1 
\int_0^{\xi-\xi_1} \!\!d\xi_l \gamma_{\bar{h}_l}(\xi_l) f(\xi-\xi_1-\xi_l) \\
\times \left[ \gamma_{g_1}(\xi_1) X_{l,0}- \gamma_{\bar{g}_1}(\xi_1) \Lambda_{l,1}\right]  \,,
\end{split}\\
\begin{split} \label{sigmas0}
 \Sigma_{0,s}^{(l)} (\xi) \equiv \int_0^\xi d\xi_1 
\int_0^{\xi-\xi_1} \!\!d\xi_l \gamma_{\bar{h}_l}(\xi_l) f(\xi-\xi_1-\xi_l) \\
\times \left[ \gamma_{h_1}(\xi_1) X_{l,0}+ \gamma_{\bar{h}_1}(\xi_1) \Lambda_{l,1}\right]\,,
\end{split}
\end{gather}
where the operators $\gamma_{g_j/\bar{g}_j}(\xi_j)$ act analogously as in \eqref{gammah}, and $f(s)=\delta(s)+\dot{p}_{\rm eN}^{(l-2)}(s)$. 

As discussed in the Sec. \ref{GMEapp}, we choose a factorized initial condition at time $t=0$, which corresponds to set $s_0 \to \infty$.
This means that in Eqs. \eqref{sigmaa0} and \eqref{sigmas0}, $X_{l,0} = R(t_{2l}-t_1)-R(t_{2l-1}-t_1)=R(\xi-\xi_1+\tau_1)-R(\xi-\xi_1+\tau_1-\tau_l)$. Moreover, $\Lambda_{l,1} =S(\xi-\xi_1+\tau_1)+S(\xi-\xi_1-\tau_l)-S(\xi-\xi_1)-S(\xi-\xi_1+\tau_1-\tau_l)$.

As before, it is more convenient to work now in the Laplace space, where many terms factorize.
In fact, one can identify (see e.g. Fig. \ref{Pwiba}) products of \textit{one} or several irreducible \textit{extended}-NIBA kernels ($\hat{F}(\lambda)$, $\hat{A_0}(\lambda)$ and $\hat{F_0}(\lambda)$) with one of the irreducible kernels containing inter-blip and blip-sojourns interactions ($\hat{\Sigma}_s^{(n)}(\lambda)$, $\hat{\Sigma}_{0,a}^{(n)}(\lambda)$ and $\hat{\Sigma}_{0,s}^{(n)}(\lambda)$).
After exchanging the integration order as done in Eq. \eqref{intd}, the probability difference reads in the Laplace space
\begin{equation} \label{pwiba2}
\begin{split}
 \hspace{-0.15cm} \hat{P}_{\rm WIBA}(&\lambda)=\hat{P}_{\rm eN}(\lambda)\\
+&\dfrac{1}{\lambda} \sum_{n=2}^\infty \Bigg\{ \sum_{l=2}^n  \left[ \prod_{j=l+1}^n -\hat{F}(\lambda) \right] \left[\hat{\Sigma}_{0,a}+\hat{\Sigma}_{0,s} \right]^{(l)} \\
+&\sum_{l=3}^n \sum_{k=2}^{l-1} \left[ \prod_{j=l+1}^n -\hat{F}(\lambda) \right] \left[\hat{\Sigma}_s \right]^{(l-k+1)}\\
&\times \left[ \prod_{j=2}^{k-1} -\hat{F}(\lambda) \right] \left[-\left(\hat{F_0}(\lambda)+\hat{A_0}(\lambda)\right)\right] 
 \Bigg\}\,.
\end{split}
\end{equation}


After some changes of variables, Eq. \eqref{pwiba2} can be recast in the form 
\begin{equation} \label{pwiba3}
\begin{split}
 \hat{P}& _{\rm WIBA}(\lambda)=\hat{P}_{\rm eN}(\lambda)\\ 
+&\dfrac{1}{\lambda} \sum_{m=0}^\infty \sum_{\gamma=0}^m \Bigg\{ \left[\hat{\Sigma}_{0,a}+\hat{\Sigma}_{0,s} \right]^{(\gamma+2)} \left[ -\hat{F}(\lambda) \right]^{m-\gamma}\\
+&\left[\hat{\Sigma}_s \right]^{(\gamma+2)} 
\left[ -\hat{F}(\lambda) \right]^{m-\gamma} (m-\gamma+1) \\
&\times\left[-\left(\hat{F_0}(\lambda) +\hat{A_0}(\lambda)\right)\right]
 \Bigg\}\,,
\end{split}
\end{equation}
which becomes, after noticing that $\sum_{m=0}^\infty \sum_{\gamma=0}^m = \sum_{\gamma=0}^\infty \sum_{m=\gamma}^\infty$: 
\begin{equation} \label{pwiba4}
\begin{split}
 \hat{P}& _{\rm WIBA}(\lambda)=\hat{P}_{\rm eN}(\lambda)\\ 
+&\dfrac{1}{\lambda} \sum_{m=0}^\infty \Bigg\{ \left[\hat{\Sigma}_{0,a}+\hat{\Sigma}_{0,s} \right]^{(m+2)} \sum_{p=0}^\infty \left[ -\hat{F}(\lambda) \right]^{p}\\
+&\left[\hat{\Sigma}_s \right]^{(m+2)} \sum_{p=0}^\infty \left[ -\hat{F}(\lambda) \right]^{p} (p+1) 
\left[-\left(\hat{F_0}(\lambda) +\hat{A_0}(\lambda)\right)\right]
 \!\Bigg\}.
\end{split}
\end{equation}
It is convenient to introduce the irreducible kernels $\hat{\Sigma}_s\equiv \sum_{m=0}^\infty \hat{\Sigma}_s^{(m+2)}$ and $\hat{\Sigma}_{0,a/s}\equiv \sum_{m=0}^\infty \hat{\Sigma}_{0,a/s}^{(m+2)}$ obtained upon summing over the number of tunneling transitions. Thus, recalling the \textit{extended}-NIBA expression \eqref{pniba2}, we find 
\begin{equation} \label{pwibaf}
\begin{split}
 \hat{P}_{\rm WIBA}(&\lambda)=\dfrac{1}{\lambda} +\dfrac{1}{\lambda} \left[-\left(\hat{F_0}(\lambda)+\hat{A_0}(\lambda)\right)\right] \\ 
&\times \sum_{m=0}^\infty \left[ -\hat{F}(\lambda) \right]^{m-1}\!\left[-\hat{F}(\lambda) +m\hat{\Sigma}_s(\lambda)\!\right]\\
+&\dfrac{1}{\lambda} \left[\hat{\Sigma}_{0,a}(\lambda)+\hat{\Sigma}_{0,s}(\lambda) \right] \sum_{m=0}^\infty \left[ -\hat{F}(\lambda) \right]^{m}.
\end{split}
\end{equation}

Keeping in mind that we should always retain the interactions $\Lambda$ and $X$ up to the first order, 
Eq. \eqref{pwibaf} can be rearranged in a more compact form as 
\begin{equation} \label{pwibafa}
\begin{split}
 \hspace{-0.3cm} \hat{P}_{\rm WIBA}(&\lambda)\approx \dfrac{1}{\lambda} +\dfrac{1}{\lambda} \left[-\left(\hat{F_0}(\lambda)+\hat{A_0}(\lambda)\right)\right. \\
&+\left.
\hat{\Sigma}_{0,a}(\lambda) +\hat{\Sigma}_{0,s}(\lambda)\right]
 \sum_{m=0}^\infty \left[ -\hat{F}(\lambda) +\hat{\Sigma}_s(\lambda)\right]^m
,
\end{split}
\end{equation}
yielding
\begin{align} \label{pwibafd}
 & \hat{P}_{\rm WIBA}(\lambda)\notag\\ 
&\hspace{-0.3cm}=\dfrac{1}{\lambda} \Bigg[1+\dfrac{-\left(\hat{F_0}(\lambda)+\hat{A_0}(\lambda)\right)+
\hat{\Sigma}_{0,a}(\lambda)+\hat{\Sigma}_{0,s}(\lambda)}{1+\hat{F}(\lambda) -\hat{\Sigma}_s(\lambda)}\Bigg]
\\
\label{pwibafd2}
 &\hspace{0.5cm}=\dfrac{1+\hat{W}_{\rm WIBA}/\lambda-\hat{K}_{\rm WIBA}^a/\lambda}{\lambda+\hat{K}_{\rm WIBA}^s},
\end{align}
where we introduced the functions 
\begin{subequations} \label{kwibalam}
 \begin{align}
  \hat{K}^{s}_{\rm WIBA}(\lambda) 
	&=\lambda \left[ \hat{F}(\lambda)-\hat{\Sigma}_{s}(\lambda) \right] \,,\\
  \hat{K}^{a}_{\rm WIBA}(\lambda) 
  	&=\lambda \left[ \hat{A_0}(\lambda)-\hat{\Sigma}_{0,a}(\lambda) \right] \,,\\
  {\hat{K}}^s_{\rm 0,WIBA}(\lambda) 
	&=\lambda \left[ \hat{F_0}(\lambda)-\hat{\Sigma}_{0,s}(\lambda) \right] \,,\\
 \begin{split}
  \hat{W}_{\rm WIBA}(\lambda) & = \hat{K}^{s}_{\rm WIBA}(\lambda)- {\hat{K}}^s_{\rm 0,WIBA}(\lambda) \,. 
 \end{split}
\end{align}
 \end{subequations}

\subsection{Generalized master equation (GME) for the WIBA model} \label{GMEw}

After multiplying both sides of Eq. \eqref{pwibafd2} by $\lambda+\hat{K}_{\rm WIBA}^s$, one can recognize the Laplace transform of  the generalized master equation Eq. \eqref{GME1}, where from Eq. \eqref{kwibalam} 
the WIBA kernels in the time domain are 
\begin{equation} \label{kwiba}
 \begin{split}
  K^{s}_{\rm WIBA}(t) &\equiv \dot{F}(t)-\dot{\Sigma}_s(t)\,,\\
  K^{s}_{0,\rm WIBA}(t) &\equiv \dot{F_0}(t)-\dot{\Sigma}_{0,s}(t)\,,\\
  K^{a}_{\rm WIBA}(t) &\equiv \dot{A_0}(t)-\dot{\Sigma}_{0,a}(t)\,,\\
  W_{\rm WIBA}(t) &\equiv K^{s}_{\rm WIBA}(t)-K^{s}_{0,\rm WIBA}(t)\,.
 \end{split}
\end{equation}

%

\begin{figure}[t]
\centering
\includegraphics[width=0.475\textwidth,bb=105 650 423 753,clip=true]{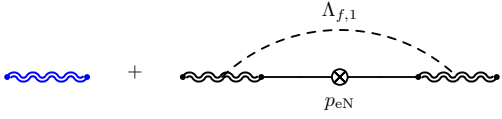}
\vspace{-0.3cm}\caption{Irreducible symmetric WIBA kernel. Notice that the lowest order reproduces the \textit{extended}-NIBA case, see Fig. \ref{Fig.interactions}b. The inner bubble represents the contribution coming from the sum of all orders in $\Delta^2$ which coincides with the symmetric part of the solution of the master equation for $P(t)$ in the \textit{extended}-NIBA case.} \vspace{-0.25cm} \label{feynwiba}
\end{figure}

The prescription \eqref{kwiba} allows the explicit evaluation of the irreducible kernels. This procedure is illustrated in App. \ref{k.wiba}. Explicitly, 
the antisymmetric WIBA kernel reads
\begin{gather}
\begin{split} \label{sigmaa0ft}
 \hspace{-0.5cm} {K_{\rm WIBA}^a}(\xi)	&= K_{\rm eN}^a(\xi) \\
&-\int_0^\xi d\xi_1 
\int_0^{\xi-\xi_1} \!\!d\xi_f\, \bar{h}_{\rm eN}(\xi_f) \,p_{\rm eN}(\xi-\xi_1-\xi_f)  \\
&\times \left[ g_{\rm N}(\xi_1) X_{f,0}- \bar{g}_{\rm N}(\xi_1) \Lambda_{f,1}\right]  
\,,
\end{split}
\end{gather}
where, as in the \textit{extended}-NIBA case, cf. \eqref{Rtilde}, corrections of order $\ddot{R}(\xi_f)\xi_f^2$ have been neglected. The functions $g_{\rm eN}$, $h_{\rm eN}$ have been defined in \eqref{ghtilde}.
Analogously, we also introduced here the functions 
\begin{equation} \label{ghtildebar}
\begin{split}
\bar{g}_{\rm eN}(\xi_j)&=\Delta^2 {\rm e}^{-S(\xi_j)} \cos(\varepsilon \xi_j) \sin[\tilde{R}(\xi_j)]\;,\\
\bar{h}_{\rm eN}(\xi_j)&=\Delta^2 {\rm e}^{-S(\xi_j)} \sin(\varepsilon \xi_j) \cos[\tilde{R}(\xi_j)]\;,
\end{split}
\end{equation}
with $\tilde{R}(t) \equiv R(t)-t \dot{R}(t)$. 
The symmetric WIBA kernel becomes 
\begin{gather}
\begin{split} \label{sigmasft}
 \hspace{-0.2cm} {K_{\rm WIBA}^s}(\xi)	&=  K_{\rm eN}^s(\xi) \\
&-\int_0^\xi d\xi_1 
\int_0^{\xi-\xi_1} \!\!d\xi_f\, \bar{h}_{\rm eN}(\xi_f) \,p_{\rm eN}(\xi-\xi_1-\xi_f)  \\
&\times \left[ \bar{h}_{\rm eN}(\xi_1) \Lambda_{f,1}\right] 
\,.
\end{split}
\end{gather}
The irreducible symmetric kernel is shown in Fig. \ref{feynwiba}.
Moreover, the function $W_{\rm WIBA}$ is now given by
\begin{equation} \label{wft}
 W_{\rm WIBA}(\xi)\equiv {K_{\rm WIBA}^s}(\xi)-K^s_{0,\rm WIBA}(\xi)\,,
\end{equation}
where
\begin{gather}
\begin{split} \label{sigmas0ft}
 \hspace{-0.2cm} K^s_{0,\rm WIBA}(\xi)	&= K^s_{0,\rm eN}(\xi) \\
&- \int_0^\xi d\xi_1 
\int_0^{\xi-\xi_1} \!\!d\xi_f\, \bar{h}_{\rm eN}(\xi_f) \,p_{\rm eN}(\xi-\xi_1-\xi_f)  \\
&\times \left[ h_{\rm N}(\xi_1) X_{f,0}+ \bar{h}_{\rm N}(\xi_1) \Lambda_{f,1}\right] 
\,.
\end{split}
\end{gather}
 
In the Eqs. \eqref{sigmaa0ft} and \eqref{sigmas0ft}, $X_{f,0}=R(\xi)-R(\xi-\xi_f)$, and for all the three kernels  $\Lambda_{f,1}=S(\xi)-S(\xi-\xi_1)+S(\xi-\xi_1-\xi_f)-S(\xi-\xi_f)$.

\subsection{Dynamics within the WIBA}
Let us now come to the WIBA predictions.
In this Section we will be showing the comparison of all the theories previously discussed, namely the NIBA, its extended version, the weak-coupling approximation (WCA) and our WIBA, for different choices of the parameters. 
In the following subsections we examine the behavior of $P(t)$ in the case of Ohmic bath ($s=1$ in the spectral density \eqref{deltas}), with cutoff frequency $\omega_c=50\Delta$, and of super-Ohmic dissipative environment (we choose $s=3$), for different choices of the frequencies $\omega_c$ and $\omega_{\rm ph}$.

\subsubsection{Ohmic case ($s=1$)}

We expect WIBA to work particularly well in the Ohmic case due to $S(t)\sim t$ at long times. This has the simultaneous effect of suppressing the blip lengths and to yield a vanishing interblip interaction at large blip separation. 

\begin{figure}[ht!]
\centering
\includegraphics[width=0.475\textwidth,bb=37 56 718 527,clip=true]{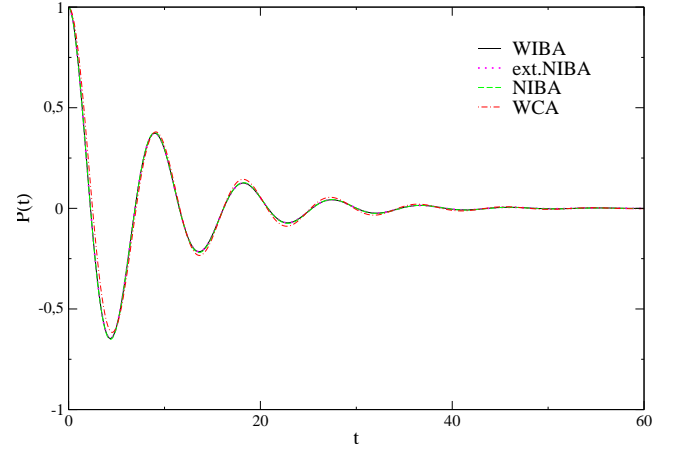}
\vspace{-0.3cm}\caption{Time evolution of $P(t)$ for an Ohmic symmetric two-state system. The WIBA (full lines) coincides with the NIBA (dashed lines) and the \textit{extended}-NIBA (dotted lines) over the whole range of parameters. Here $\alpha=0.1$ and $T=0.1$ (in units of $\Delta$) have been chosen. For such a choice, the WCA (dot-dashed lines) slightly deviates from all other predictions.} \vspace{-0.cm} \label{wiba0}
\end{figure}

We first present (Fig. \ref{wiba0}) the results for an Ohmic symmetric two-state system ($\varepsilon=0$) for a generic choice of the parameters. For the symmetric case, the NIBA is expected to predict the correct time evolution over the whole range of coupling strengths and temperatures. \begin{figure}[hb!]
\centering
\includegraphics[width=0.475\textwidth,bb=37 56 718 527,clip=true]{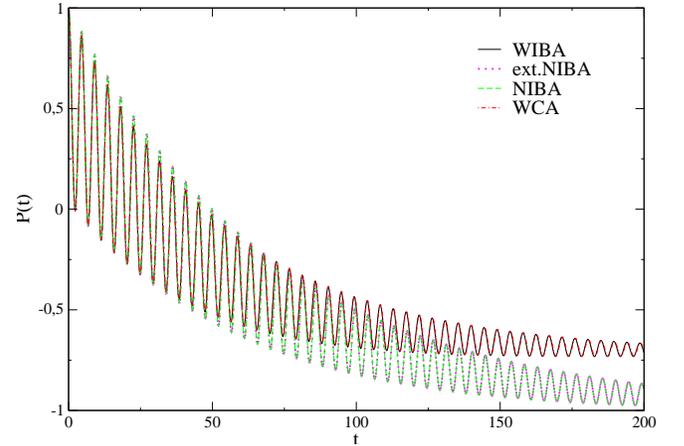}
\vspace{-0.3cm}\caption{Time evolution of $P(t)$ for Ohmic damping and finite bias. The chosen parameters are $\alpha=0.01$, $T=0.1$ and $\varepsilon=1$ (in units of $\Delta$). The WIBA (full lines) coincides with the WCA (dot-dashed lines) for  low temperatures and small coupling, while the NIBA (dashed lines) and the \textit{extended}-NIBA (dotted lines) predict an unphysical asymptote, for an asymmetric system.} \vspace{-0.cm} \label{wiba0.01}
\end{figure}
One sees a complete agreement of the WIBA with the (\textit{extended}-) NIBA. For the chosen parameters, also WCA well agrees with the WIBA predictions. 

The situation becomes more intricate for the case of finite bias ($\varepsilon \ne 0$), since on the one hand NIBA is expected to fail at low temperatures and damping, while the WCA becomes inappropriate at large temperatures and/or damping. In the following, we fix the external bias as $\varepsilon=\Delta$. 

Fig. \ref{wiba0.01} shows a case of low temperature $k_{\rm B}T\le E=\sqrt{\varepsilon^2+\Delta^2}$ and small damping $\alpha$, namely $\alpha=0.01$ and $T=0.1$ (in units of $\Delta$). 
One can see that the agreement between WIBA and WCA is striking, whereas both NIBA and \textit{extended}-NIBA fail to reach the correct asymptotic value, predicting an unphysical symmetry breaking.

\begin{figure}[ht]
\centering
\includegraphics[width=0.475\textwidth,bb=37 56 718 527,clip=true]{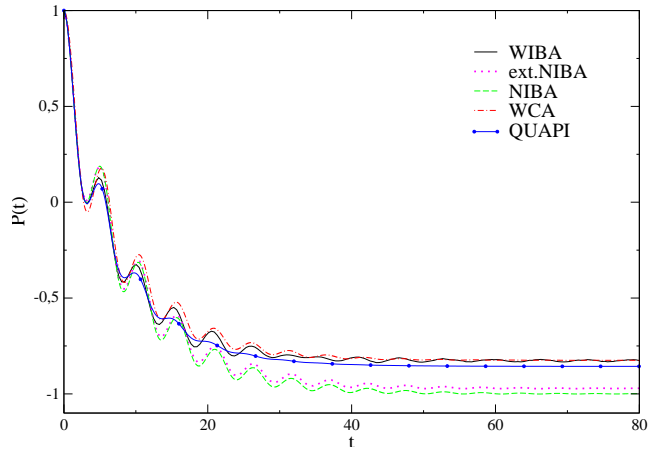}
\vspace{-0.3cm}\caption{Time evolution of $P(t)$ for $\alpha=0.1$, $T=0.1$. The WIBA (full lines) exhibits more coherence than the QUAPI (lines with bullets), being however its predictions closer to the QUAPI than the WCA (dot-dashed lines). The \textit{extended}-NIBA (dotted lines) is also getting closer to the QUAPI predictions, the NIBA (dashed lines) still predicting a strong localization in the left well.} \vspace{-0.3cm} \label{wiba0.1}
\end{figure}

In Figs.\ \ref{wiba0.1}, \ref{wiba0.25,0.01} and  \ref{wiba0.25,0.1}, a comparison between WIBA and the numerical \textit{ab-initio} path-integral approach (QUAPI \cite{Makarov95}) is made, in order to proof the validity of our theory in a regime of intermediate-to-high  temperatures and coupling strength. 
As the coupling strength is raised to $\alpha=0.1$ (Fig. \ref{wiba0.1}), while keeping the temperature constant, one sees that the asymptote of the \textit{extended}-NIBA is slowly moving from the NIBA one towards the QUAPI one. At the same time, the WCA also disagrees with the WIBA predictions, the former starting to be invalid for higher coupling strength. 
The WIBA, however, shows the presence of some unphysical ``beatings'' at the onset of the long-time dynamics, whose origin is not yet well understood. 
Hence, some more investigations seem to be required, in order to better understand the theory in this regime of parameters.

\begin{figure}[ht!]
\centering
\includegraphics[width=0.475\textwidth,bb=37 56 718 527,clip=true]{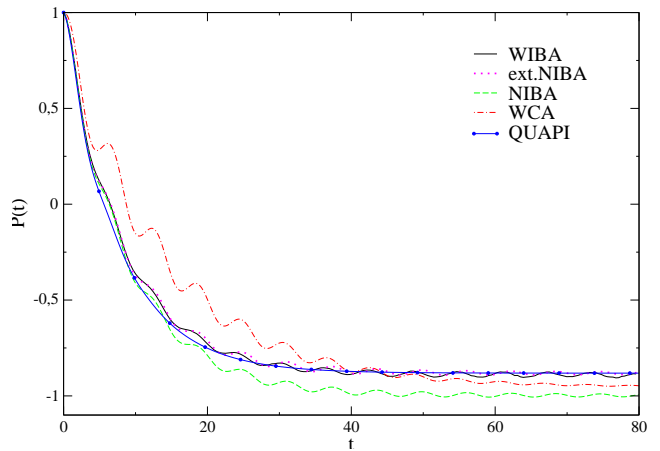}
\vspace{-0.3cm}\caption{Time evolution of $P(t)$ for $\alpha=0.25$, $T=0.01$. The WCA (dot-dashed lines) completely fails in describing the short-time dynamics and also predicts a wrong asymptote. The WIBA (full lines) as well as the \textit{extended}-NIBA~(dotted lines) very well agree with the QUAPI predictions (lines with bullets), despite some spurious oscillations, still contained in the theory. The disagreement with the NIBA is striking, already at this intermediate-to-high regime of parameters, although the NIBA correctly describes the~short-time~dynamics.} \vspace{-0.cm} \label{wiba0.25,0.01}
\end{figure}

As one raises the coupling strength ($\alpha=0.25$) (Fig. \ref{wiba0.25,0.01}), by lowering the temperature ($T=0.01$), one can observe that the WCA completely fails in predicting the right asymptote or the intermediate dynamics. The NIBA also predicts a wrong asymptotic value, while correctly describing the short-time dynamics. For such a choice of parameters, the \textit{extended}-NIBA very well agrees with QUAPI, although some oscillations are still present in the theory. The WIBA also smoothly oscillates close to the QUAPI predictions.

\begin{figure}[hb!]
\centering
\includegraphics[width=0.475\textwidth,bb=37 56 718 527,clip=true]{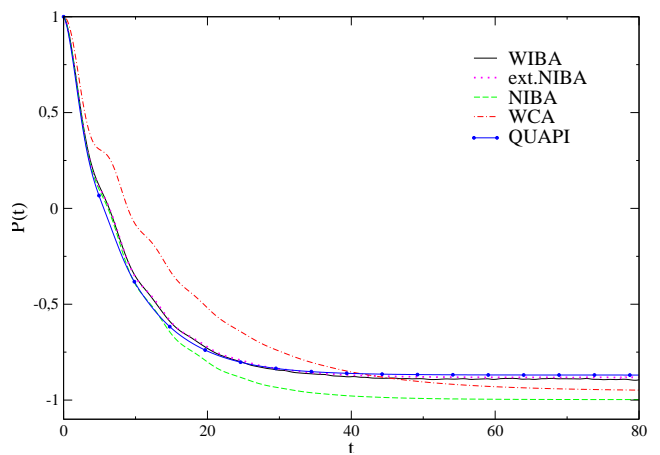}
\vspace{-0.3cm}\caption{Time evolution of $P(t)$ for $\alpha=0.25$, $T=0.1$. In this regime the \textit{extended}-NIBA (dotted lines) as well as the WIBA (full lines) are almost indistinguishable from QUAPI (lines with bullets), whereas both WCA (dot-dashed lines) and NIBA (dashed lines) fail in describing the dynamics, the latter being, however, able to reproduce the short-time dynamics.} \vspace{-0.2cm} \label{wiba0.25,0.1}
\end{figure}

Finally, Fig. \ref{wiba0.25,0.1} shows that, for the parameters $\alpha=0.25$, $T=0.1$, the WIBA and the \textit{extended}-NIBA are almost indistinguishable from the QUAPI predictions. The WCA and the NIBA fail to describe the dynamics, even though the NIBA correctly reproduces the short-time dynamics. 
By looking at Figs.\ \ref{wiba0.1}, \ref{wiba0.25,0.01} and \ref{wiba0.25,0.1}, it is interesting to notice that the WIBA predicts more coherence than QUAPI. The overall agreement with QUAPI though remains very good.

Hence, we showed that WIBA works very well for strong damping and/or high temperatures and small coupling strength and/or low temperatures, where is able to reproduce previous perturbative theories. In an intermediate regime of parameters, WIBA is shown to give an appropriate description of the dynamics.

\subsubsection{Super-Ohmic case ($s=3$)}

Let us now consider the predictions of the WIBA
in presence of a super-Ohmic bath. We choose to show here a very common situation, corresponding to $s=3$.
In contrast to the Ohmic case, analyzed above, for the super-Ohmic case is unclear if the WIBA can perform well over a broad regime of parameters.
In fact, since $S(t \sim \Delta^{-1})$ differs only little from its asymptotic value $S(t \to \infty)$, the interblip interactions are weak and the WCA is expected to be a good approximation in a wide regime of parameters. This also implies that $S(t)$ is \textit{not} effective in suppressing long-blip lenghts and the NIBA might not be justified. 
Notice that we expect WIBA to work well in all the situations such that $1< s<2$, since for those parameters the function $S(t)$ increases with time. 
%

In the following figures, we keep temperature, external bias and coupling constant fixed ($T=0.1$, $\varepsilon=1$, $\delta_3=0.01$, in units of $\Delta$), while keeping the ratio $\omega_c/\omega_{\rm ph}$ constant ($\omega_c/\omega_{\rm ph}=8$). This choice is reasonable and agrees with some experiments \cite{Vorojtsov05}. 
\begin{figure}[htb!]
\centering
\includegraphics[width=0.475\textwidth,bb=37 56 718 527,clip=true]{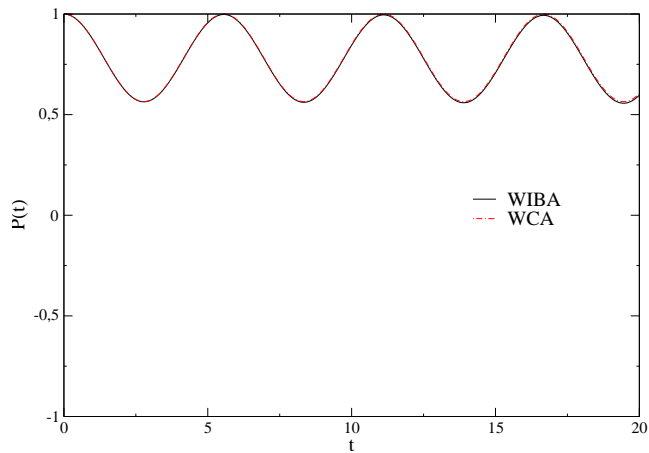}
\vspace{-0cm}\caption{Time evolution of $P(t)$ for $\delta_s=0.01$, $\omega_c=200$ ($T=0.1$, $\varepsilon=1$). Here one sees no difference between WCA (dot-dashed lines) and WIBA (full lines), showing that the chosen parameters lie in an effective weak-coupling regime.} \vspace{-0.2cm} \label{wiba3-25}
\end{figure}
The ratio $\omega_c/\omega_{\rm ph}$ acts as an effective correction factor to the real coupling strength, see Eqs. \eqref{S} and \eqref{R}. 
Moreover, since we fix the temperature, in the bath correlation function corrections proportional to the temperature ($\kappa=1/\hbar\beta\omega_c$) become important as the cutoff frequency is decreased.

In Fig.\  \ref{wiba3-25} the cutoff frequency $\omega_c=200$ is assumed to be very large compared to the system frequency scales. There, one sees that the WIBA well matches the WCA predictions, as expected in the regime of weak damping.

\begin{figure}[hbt!]
\centering
\includegraphics[width=0.475\textwidth,bb=37 56 718 527,clip=true]{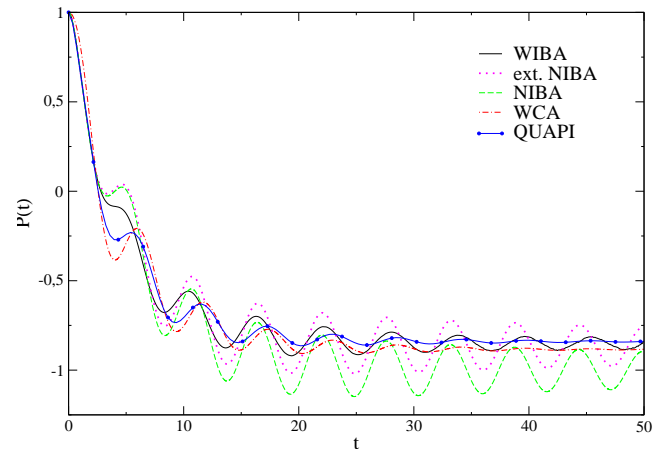}
\vspace{-0.cm}\caption{Time evolution of $P(t)$ for $\delta_s=0.01$, $\omega_c=1$ ($T=0.1$, $\varepsilon=1$). Here we notice discrepancies between WIBA (full lines) and QUAPI (full lines with bullets), since the \textit{extended}-NIBA predictions (dotted lines) also account for too many oscillations. Nevertheless, it gives much better results than NIBA (dashed lines), predicting unphysical values for $P(t)$. Finally, the WCA (dot-dashed lines) is so far the best model in this regime, although it lies apart as well from the QUAPI predictions, since the perturbative approach for such parameters begins to fail.} \vspace{-0.cm} \label{wiba3-1}
\end{figure}

On the other hand, Fig.\  \ref{wiba3-1} shows the WIBA predictions for $P(t)$ when the cutoff frequency $\omega_c$ is of the order of the tunneling frequency $\Delta$, namely when the bath becomes ``slow'', a case where the WCA and the NIBA are expected to fail. 
This case is the most difficult one, since the bath is very coherent and memory effects are to be taken into account, which requires to perform a very good description of the full bath dynamics. 
One sees that the NIBA completely fails to reproduce the dynamics, even reaching unphysical values. The \textit{extended}-NIBA works better, approaching closer the QUAPI predictions. Nevertheless, too few correlations are taken into account, and it oscillates still too much with respect to the numerical plot of QUAPI. The WIBA shows discrepancies from the QUAPI as well, being still ``too'' coherent, 
even though its predictions are  more accurate than the \textit{extended}-NIBA.
 The WCA, despite better than WIBA in this regime, also lie apart from the numerical prediction of QUAPI. In this range of parameters, the temperature-dependent corrections in the bath correlation function $Q$ become relevant and the perturbative weak-coupling approach begins to fail. 
Thus, more analysis of the complicated super-Ohmic case is to be done, in order to better understand the different dynamical situations which take place by varying the coupling strength $\delta_s$, the cutoff frequency $\omega_c$ and the phonon frequency $\omega_{\rm ph}$.

\section{Conclusions}

To conclude, we presented here a novel analytical scheme dubbed WIBA (Weakly-Interacting Blip Approximation) which is able to match between diverse approximation schemes for different parameters choices.
In particular, the WIBA is valid over a large regime of temperatures and coupling strength for a wide class of spectral densities \eqref{deltas}. The WIBA reproduces very well weak-coupling approximation schemes for small damping and temperature and the well-known NIBA in the opposite regime of strong damping and high temperature. 
It yields a good, though not perfect agreement, with \textit{ab-initio} QUAPI calculation in the regime of intermediate temperatures and damping.
Hence, the WIBA overcomes the dim validity of the perturbative approaches discussed in the previous Sections which, up to now, was only possible via numerical \textit{ab-initio} schemes.
It also serves as benchmark to limiting simple analytical schemes, in order to proof their range of applicability over the various regimes, and combine weak-coupling and strong-damping approaches in one single, unified model.

Our approach is based on the consideration that bath-induced blip-blip and blip-sojourn interactions are weak in the whole regime of parameters for the spectral densities \eqref{deltas}. It is expecially needed in several contexts, and we mention here few of them: i) 
The WIBA could be very useful when experimentally the bath temperature or the external bias are varied over a wide range: In fact, WCA and NIBA are unreliable at high temperatures and low-to-intermediate driving, respectively.
ii) It should be used to investigate the situation of several TLS's interacting with a common heat bath \cite{Stock-glass}, as in glasses \cite{Würger}, at low temperature being characterized by a broad distribution of tunneling parameters and asymmetries.

We must, however, notice, that in the case of ``slow'' environments with cut-off frequency of the order of the tunneling frequency, still our model has to be improved, since neither the WIBA nor other analytical approximation schemes are able to reproduce the correct onset of decoherence which in fact takes place.
 This situation occurs e.g. in  non-adiabatic electron transfer \cite{Garg85} or for charge qubits interacting with piezo-electric phonons \cite{Vorojtsov05}.


\section*{Acknowledgments}

We wish to thank A. Donarini for helpful discussions and L. Hartmann, D. Bercioux, M. Storcz for useful hints on numerics.
Support under the DFG programs GKR638 and SFB631 is acknowledged. We also acknowledge support within the projects EU-EuroSQIP (IST-3-015708-IP) and MIUR-PRIN 2005 (200502 2977).


\appendix

\section{Derivation of the NIBA kernels}
\label{k.niba}

In the non-interacting blip approximation, as mentioned in the Sec. \ref{eniba},  one neglects all blip-blip correlations and blip-sojourns interactions in Eq. \eqref{influencei}, i.e. $\Phi_{\rm inter}^{(n)}=0$. The blip-preceeding-sojourn interactions are neglected as well ($\Phi_{\rm bps}^{(n)}= 0$), i.e. one sets $X_{j,j-1} \approx R_{2j,2j-1}$. As a consequence, the influence function (\ref{influence}) factorizes into
individual  influence factors depending only on the dipole length $\tau_j:=t_{2j}-t_{2j-1}$.
Following an analogous procedure as in Sec. \ref{GMElam}, the NIBA series expression can be derived, and the corresponding GME, of the same form as in Eq. \eqref{GMEen}, is obtained.
The kernels finally read
\begin{subequations} \label{kniba2}
 \begin{align}
  K^{s}_{\rm N}(t) &=\Delta^2  {\rm e}^{-S(t)} \cos(\varepsilon t) \cos[R(t)]\,,\\
  K^{a}_{\rm N}(t) &  =\Delta^2  {\rm e}^{-S(t)} \sin(\varepsilon t) \sin[R(t)]\,,\\
  K^{s}_{\rm 0,N}(t) & =\Delta^2  {\rm e}^{-S(t)} \cos(\varepsilon t) \cos[R(t)]\,,\\
  W_{\rm N}(t) &\equiv K^{s}_{\rm N}(t)-K^{s}_{\rm 0,N}(t) = 0\,.
 \end{align}
\end{subequations}

 The kernels are of lowest order in the tunneling matrix $\Delta$
 but are {\em non}-perturbative in $\delta_s$. The NIBA is expected to be a good approximation whenever the average time spent in a blip is much larger than the time spent in a sojourn.

In general, the inter-dipole interaction can be safely neglected
for all temperatures for sub-Ohmic damping $s<1$, and the TLS is
expected to exhibit incoherent dynamics even for very small coupling
$\delta_s$.
For Ohmic and super-Ohmic
damping, NIBA is expected to
be a good approximation at high enough temperature and/or strong
damping. In particular, spectral densities of the Ohmic form reach
for large temperatures faster the asymptotic behavior $S(t)\propto
t$ at long times, implying that the correlations in $\Lambda_{j,k}$
cancel out exactly.  


\section{Evaluation of the WIBA kernels} \label{k.wiba}

In this Appendix, the prescription \eqref{kwiba} to evaluate the WIBA kernels in the time domain is illustrated.
Let us consider, to fix the ideas, $\Sigma_{0,a}(\xi)=\sum_l \Sigma_{0,a}^{(l)}(\xi)$. 
The kernel $\Sigma_{0,a}^{(l)}$, given in Eq. \eqref{sigmaa0}, is reported here for clarity:
\begin{gather}
\begin{split}
 \Sigma_{0,a}^{(l)} (\xi) \equiv \int_0^\xi d\xi_1 
\int_0^{\xi-\xi_1} \!\!d\xi_l \gamma_{\bar{h}_{l}}(\xi_l) f(\xi-\xi_1-\xi_l) \\
\times \left[ \gamma_{{g}_1}(\xi_1) X_{l,0}- \gamma_{\bar{g}_1}(\xi_1) \Lambda_{l,1}\right]  \,,
\end{split}
\end{gather}
with $X_{l,0}=R(\xi-\xi_1+\tau_1)-R(\xi-\xi_1+\tau_1-\tau_l)$. Notice that the operators $\gamma$'s depend on the functions $g/\bar{g}$ ($h/\bar{h}$) defined in Eqs. \eqref{ghbwiba} and \eqref{kniba0n}. 

 In the function $f(s)$ which appears in the expression for the kernel (see Eq. \eqref{fs}), the first derivative of the conditional probability $\dot{p}_{\rm eN}^{(l-2)}(s)$ is present. Then, it could be integrated out by parts, obtaining 
\begin{gather}
\begin{split} \label{sigmaa0i}
 \Sigma_{0,a}^{(l)} &(\xi) = \int_0^\xi d\xi_1 
\int_0^{\xi-\xi_1} \!\!d\xi_l \, \dot{\gamma}_{\bar{h}_{l}}(\xi_l) \\
&\times p_{\rm eN}^{(l-2)}(\xi-\xi_1-\xi_l) \left[ \gamma_{{g}_1}(\xi_1) X_{l,0}- \gamma_{\bar{g}_1}(\xi_1) \Lambda_{l,1}\right]  
\,.
\end{split}
\end{gather}
Here, the first derivative of the operator $\gamma_{\bar{h}_{l}}(\xi_l)$ acts on the functions  $X_{l,0}=R(\xi-\xi_1+\tau_1)-R(\xi-\xi_1+\tau_1-\tau_l)$ and $\Lambda_{l,1}=S(\xi-\xi_1+\tau_1)+S(\xi-\xi_1-\tau_l)-S(\xi-\xi_1)-S(\xi-\xi_1+\tau_1-\tau_l)$ as
\begin{gather} \label{gammahl}
\begin{split} 
 &{\phantom =}\dot{\gamma}_{\bar{h}}(\xi_l) X_{l,0} \\
&= \partial_{\xi_l} \int_0^{\xi_l} \!\!d\tau_l\, \bar{h}(\tau_l,\xi_l-\tau_l) X_{l,0} \\
&\approx \partial_{\xi_l} \int_0^{\xi_l} \!\!d\tau_l\, \bar{\bar{h}}(\tau_l,\xi_l) X_{l,0} \\
&= \bar{h}_{\rm eN}(\xi_l) \left[R(\xi-\xi_1+\tau_1)-R(\xi-\xi_1+\tau_1-\xi_l) \right], 
\end{split}
\end{gather}
and
\begin{gather}
\begin{split} 
 &{\phantom =}\dot{\gamma}_{\bar{h}}(\xi_l) \Lambda_{l,1} \\
&= \partial_{\xi_l} \int_0^{\xi_l} \!\!d\tau_l\, \bar{h}(\tau_l,\xi_l-\tau_l) \Lambda_{l,1} \\
&\approx \partial_{\xi_l} \int_0^{\xi_l} \!\!d\tau_l\, \bar{\bar{h}}(\tau_l,\xi_l) \Lambda_{l,1} \\
&= \bar{h}_{\rm eN}(\xi_l) \left[ S(\xi-\xi_1+\tau_1)+S(\xi-\xi_1-\xi_l)\right.\\
&\left.-S(\xi-\xi_1)-S(\xi-\xi_1+\tau_1-\xi_l) \right]\,. 
\end{split}
\end{gather}

As already observed when calculating the \textit{extended}-NIBA kernels, the explicit dependence of the integrand of Eq. \eqref{gammahl} on $\xi_l$ yields an integral form for $\dot{\gamma}_{\bar{h}}(\xi_l) X_{l,0}$. 
An approximate form is obtained if we apply the same prescription \eqref{Rtilde} as for the \textit{extended}-NIBA. Namely, one approximates the bath correlation differences in $Y_{j,j-1}$ with the first derivative of $R(t)$. Hence, for $g$ and $h$ Eq. \eqref{gheniba} holds, whereas for $\bar{g}$ and $\bar{h}$ it follows ($l>1$) 
\begin{equation} \label{ghenibap}
\begin{split}
\bar{g}(\tau_l, \xi_l-\tau_l)&\approx \bar{\bar{g}}(\tau_l, \xi_l)\\
&=\Delta^2 {\rm e}^{-S(\tau_l)} \cos(\varepsilon \tau_l) \sin[R(\tau_l)-\tau_l \dot{R}(\xi_l)]\;,\\
\bar{h}(\tau_l, \xi_l-\tau_l)&\approx \bar{\bar{h}}(\tau_l, \xi_l)\\
&=\Delta^2 {\rm e}^{-S(\tau_l)} \sin(\varepsilon \tau_l) \cos[R(\tau_l)-\tau_l \dot{R}(\xi_l)]\;,
\end{split}
\end{equation}
the corrections being of order $\CMcal{O}[\tau_l^2 \ddot{R}(\xi_l)]$. We define $\bar{g}_{\rm eN}(\xi_l)\equiv \bar{\bar{g}}(\xi_l, \xi_l)$ and $\bar{h}_{\rm eN}(\xi_l)\equiv \bar{\bar{h}}(\xi_l, \xi_l)$.

In order to evaluate the antisymmetric WIBA kernel $K_{\rm WIBA}^a$ from the prescription \eqref{kwiba}, one sees that the previous expression must be derived with respect to $\xi$. 
After substituting $u\equiv \xi-\xi_1$ in the Eq. \eqref{sigmaa0i}, we get
\begin{gather}
\begin{split} \label{sigmaa0ii}
 \Sigma_{0,a}^{(l)} &(\xi) = \int_0^\xi du 
\int_0^{u} \!\!d\xi_l\, \bar{h}_{\rm eN}(\xi_l) \\
 &\times p_{\rm eN}^{(l-2)}(u-\xi_l) \left[ \gamma_{{g}_1}(\xi-u) X_{l,0}- \gamma_{\bar{g}_1}(\xi-u) \Lambda_{l,1}\right]  
,
\end{split}
\end{gather}
with $X_{l,0}=R(u+\tau_1)-R(u+\tau_1-\xi_l)$ and $\Lambda_{l,1}=S(u+\tau_1)+S(u-\xi_l)-S(u)-S(u+\tau_1-\xi_l)$.

Then, in order to get the derivative of $\Sigma_{0,a}^{(l)} (\xi)$ with respect to $\xi$, it is enough to evaluate
\begin{gather}
\begin{split} 
  \partial_{\xi} &\left[ \gamma_{{g}_1}(\xi-u) X_{l,0}- \gamma_{\bar{g}_1}(\xi-u) \Lambda_{l,1}\right] \\
=\partial_{\xi} & \int_0^{\xi-u} \!\!d\tau_1 \left[ {g}_1(\tau_1, s_0 \to \infty) X_{l,0}- \, \bar{g}_1(\tau_1, s_0 \to \infty)  \Lambda_{l,1}\right]\\
=\phantom{\partial_{\xi}} &\hspace{-0.2cm} {g}_{\rm N}(\xi-u) X_{l,0}- \, \bar{g}_{\rm N}(\xi-u)  \Lambda_{l,1} \,,
\end{split}
\end{gather}
where now $X_{l,0}=R(\xi)-R(\xi-\xi_l)$ and $\Lambda_{l,1}=S(\xi)+S(\xi-\xi_1-\xi_l)-S(\xi-\xi_1)-S(\xi-\xi_l)$.

Hence, we obtain for the $l$-th order (after the derivative, we substitute back $\xi_1\equiv \xi-u$)
\begin{gather}
\begin{split} \label{sigmaa0p}
&\partial_{\xi} \Sigma_{0,a}^{(l)} (\xi) \\
	&= \int_0^\xi d\xi_1 
\int_0^{\xi-\xi_1} \!\!d\xi_l\, \bar{h}_{\rm eN}(\xi_l) p_{\rm eN}^{(l-2)}(\xi-\xi_1-\xi_l)  \\
&\times \left[ {g}_{\rm N}(\xi_1) X_{l,0}- \bar{g}_{\rm N}(\xi_1) \Lambda_{l,1}\right]  
\,.
\end{split}
\end{gather}


In the end, the antisymmetric WIBA kernel reads
\begin{gather}
\begin{split} \label{sigmaa0f}
 K& _{\rm WIBA}^a (\xi)\equiv K_{\rm eN}^a(\xi) -\sum_{\sigma=2}^\infty {K_{\rm WIBA}^a}\!^{(\sigma)} \!(\xi) \\
	&= K_{\rm eN}^a(\xi) \\
&-\int_0^\xi d\xi_1 
\int_0^{\xi-\xi_1} \!\!d\xi_f\, \bar{h}_{\rm eN}(\xi_f) \,p_{\rm eN}(\xi-\xi_1-\xi_f)  \\
&\times \left[ g_{\rm N}(\xi_1) X_{f,0}- \bar{g}_{\rm N}(\xi_1) \Lambda_{f,1}\right]  
\,,
\end{split}
\end{gather}
where we introduced $p_{\rm eN}(t) \equiv \sum_{\sigma=2}^\infty p_{\rm eN}^{(\sigma-2)}(t)$. 
Following similar lines, we obtain for the symmetric WIBA kernels:
\begin{gather}
\begin{split} \label{sigmas0f}
 K^s& _{0,\rm WIBA}(\xi) \equiv K^s_{0,\rm eN}(\xi) -\sum_{\sigma=2}^\infty K^s_{0,\rm WIBA}\!^{(\sigma)} \!(\xi) \\
	&= K^s_{0,\rm eN}(\xi) \\
&- \int_0^\xi d\xi_1 
\int_0^{\xi-\xi_1} \!\!d\xi_f\, \bar{h}_{\rm eN}(\xi_f) \,p_{\rm eN}(\xi-\xi_1-\xi_f)  \\
&\times \left[ h_{\rm N}(\xi_1) X_{f,0}+ \bar{h}_{\rm N}(\xi_1) \Lambda_{f,1}\right]  
\,,
\end{split}
\end{gather}
and 
\begin{gather}
\begin{split}
 K& _{\rm WIBA}^s(\xi) \equiv K_{\rm eN}^s(\xi)- \sum_{\sigma=2}^\infty {K_{\rm WIBA}^s}\!^{(\sigma)} \!(\xi) \notag \\
	&=  K_{\rm eN}^s(\xi) 
\end{split}
\end{gather}
\begin{gather}
\begin{split}  \label{sigmasf}
&-\int_0^\xi d\xi_1 
\int_0^{\xi-\xi_1} \!\!d\xi_f\, \bar{h}_{\rm eN}(\xi_f) \,p_{\rm eN}(\xi-\xi_1-\xi_f)  \\
&\times \left[ \bar{h}_{\rm eN}(\xi_1) \Lambda_{f,1}\right]  
\,.
\end{split}
\end{gather}
We identified in all equations the last time interval with $\tau_f$ and correspondingly $\xi_f$. 
In this last equation, we also identified the first time interval $\tau_k$ with $\tau_1$ (therefore $\xi_k$ with $\xi_1$) and $\Lambda_{f,k}$ with $\Lambda_{f,1}$. 
 
In Eqs. \eqref{sigmaa0f} and \eqref{sigmas0f}, is $X_{f,0}=R(\xi)-R(\xi-\xi_f)$, and for all the three kernels  $\Lambda_{f,1}=S(\xi)+S(\xi-\xi_1-\xi_f)-S(\xi-\xi_1)-S(\xi-\xi_f)$.
In Eq. \eqref{sigmasf}, the term proportional to $X_{f,0}$ is absent because it vanishes after the integration by parts and the derivative, since it does not depend on $s_0$.

\mbox{~}



\begin{thebibliography}{99}
\bibitem{note} Blips/sojourns are time intervals spent in off-diagonal and diagonal elements of the reduced density matrix, respectively.  In the NIBA, the correlations among them are fully neglected.
\end{thebibliography}
\end{document}